\documentclass{aa}
\usepackage{graphicx}
\def\lesssim{\mathrel{\hbox{\rlap{\hbox{\lower4pt\hbox{$\sim$}}}\hbox{$<$}}}}
\def\gtrsim{\mathrel{\hbox{\rlap{\hbox{\lower4pt\hbox{$\sim$}}}\hbox{$>$}}}}

\begin{document}
   \title{Spectral analysis of sdB stars from the Hamburg Quasar
	 Survey\thanks{Based on Observations collected at the German-Spanish 
	Astronomical Center (DSAZ), Calar Alto, operated by the 
	Max-Planck-Institut f\"ur Astronomie Heidelberg jointly
	with the Spanish National Commission for Astronomy.
	}
	}

   \author{H. Edelmann\inst{1}, U. Heber\inst{1}, H.-J. Hagen\inst{2},
           M. Lemke\inst{1}, S. Dreizler\inst{3}, R. Napiwotzki\inst{1}, D. Engels\inst{2}
         }

   \offprints{H. Edelmann\\email: {\tt edelmann@sternwarte.uni-erlangen.de}}

   \institute{Dr. Remeis-Sternwarte, Astronomisches Institut der Universit\"at~Erlangen-N\"urnberg,
           Sternwartstr.~7,
           96049~Bamberg, Germany
	   \and
     Hamburger Sternwarte,  Gojenbergsweg~112, 21029~Hamburg, Germany
     \and
	   Institut f\"ur Astronomie und Astrophysik, Sand~1, 72076~T\"ubingen, Germany 
     }

   \date{Received 03 May 2002 / Accepted 17 January 2003}

   \abstract{
   We present the results of a spectral analysis of a large sample of sub\-dwarf B stars 
   selected from follow-up observations of candidates from the Hamburg Quasar Survey.
   Fundamental parameters (effective temperature, gravity, and helium abundance) 
   were determined by matching synthetic line profiles calculated from model
   atmospheres to all hydrogen and helium absorption lines present in the observed 
   optical spectra. 
   The derived helium abundances are compared with the atmospheric parameters to search for
   possible trends.
   We discovered a correlation between the helium abundance
   and the effective temperature: the larger the temperature, 
   the larger the photospheric helium content of sdB stars.
   Additionally, a 
   separation into two sequences of sdB stars in the effective 
   temperature -- helium abundance plane is detected.
   We compared our analysis results with data from the literature.
   The stars from our sample are found to be somewhat more luminous. This can only partly
   be explained by NLTE effects.
   Three apparently normal B stars were  discovered, which could be massive stars 
   far away from the galactic plane (7-19kpc).
   Radial velocities were measured for 23 stars from which we discovered 
   a new radial velocity variable sdB star.
   }

   \authorrunning{Edelmann et al.}
   \maketitle
%

\section{Introduction}
\begin{figure*}
\vspace{11.8cm}
\includegraphics{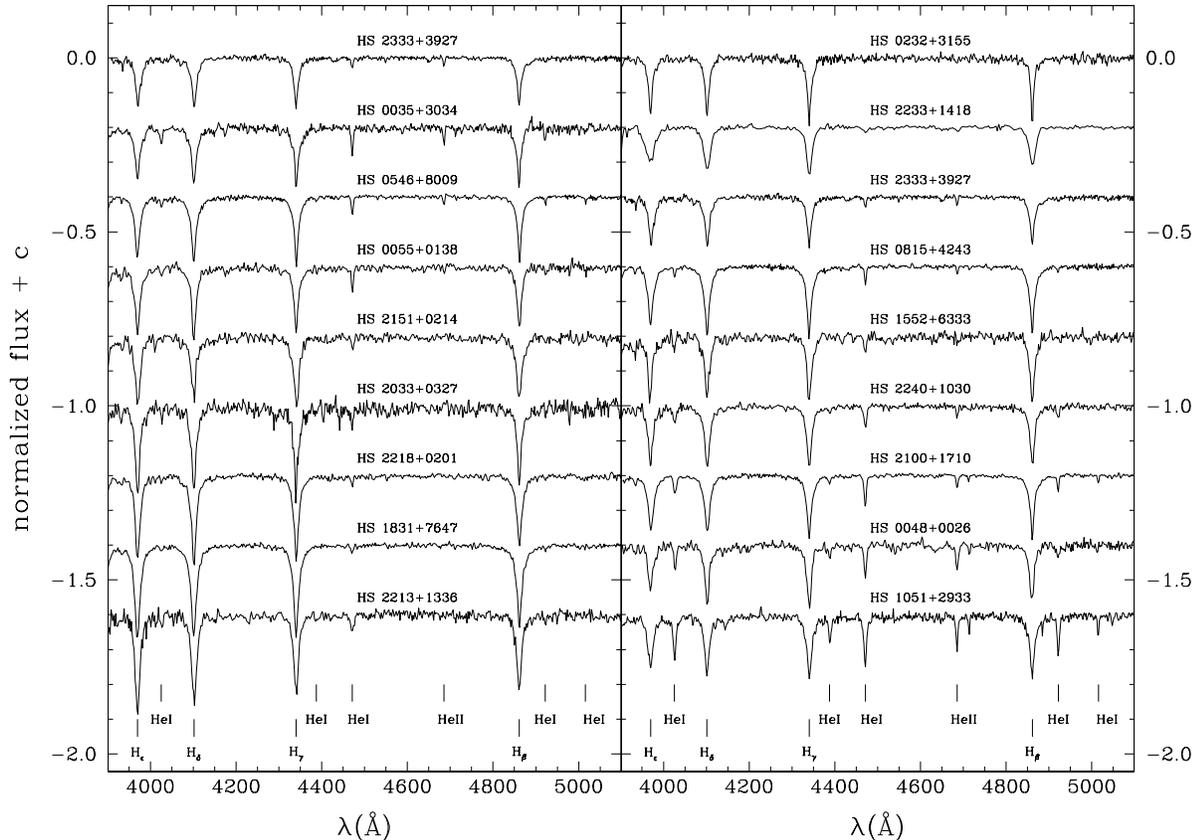}
	\caption[]{Normalized spectra of selected programme stars. Left hand panel: The spectra are
arranged in order of effective tem\-pe\-ra\-tu\-re, ranging from $\sim20\, 000$K at the bottom
to $\sim 38 \, 500$K at the top. 
The selected stars all have roughly similar gravities and helium abundances  
($\log(g) \approx 5.5\, [{\it cgs}],\; n(He)/n(H) \approx 0.01$). 
Right hand panel: Helium abundance variations: $ n(He)/n(H) < 10^{-4}$ at the top 
to $n(He)/n(H) \sim 0.25$ at the bottom.
The selected stars all have roughly similar effective temperatures and gravities 
($T_{\rm eff} \approx 35 \, 000$K$ , \; \log(g) \approx 6.0$ [{\it cgs}]).
 	\label{samplespectra}
}
\end{figure*}
The Hamburg Quasar Survey (HQS) was carried out, starting in 1980, with the 80cm Schmidt telescope
at the German-Spanish Astronomical Center (DSAZ) on Calar Alto, Spain (Hagen et al. 
\cite{Hag95}).
Although it was primarily initiated as a quasar survey, it is also a very rich source 
of faint blue stars.
Unlike the Palomar Green (PG) survey (Green et al. \cite{Gre86}) and most other surveys,
objective prism spectra (spectral resolution of 45\AA\ FWHM at H$_\gamma$) were obtained.
Afterwards, the HQS plates were digitized in Hamburg using a PDS microdensitometer.
A search software selects quasar candidates as well as faint blue 
stars from the 30\,000 -- 50\,000 spectra per plate in the
magnitude range of 13\fm 5 $\le B \le$ 18\fm 5. 
Selection criteria are blue continua and/or emission lines (Hagen et al. \cite{Hag95}).

In a collaboration between the institutes in Hamburg, Kiel, T\"ubingen, and Bamberg, follow-up 
observations and analyses of visually selected candidates of 
hot stars were performed.

The current database of follow-up observations contains 
well over 400 confirmed stars. 
The dominant fractions of the list are hot subdwarfs ($\sim 50\%
$) and white dwarfs ($\sim 30\%
$).
A lot of very rare and unusual stars also were found.
The discovery of four PG 1159, nine hot DO, and five DAO white dwarfs, so far, are 
the highlights
of the follow-up observations (Heber, Dreizler \& Hagen \cite{Heb96} and references therein). 
A comprehensive summary of the results from the HQS follow-up observations of hot stars 
can be found in Heber, Jordan \& Weidemann (\cite{HJW91}), Dreizler et al. (\cite{Dre94}), 
Lemke et al. (\cite{Lem97a}, sdO stars), and Homeier et al. (\cite{Hom98}, \cite{Hom99}, DA white dwarfs).

The present analysis focuses on the subluminous B, or {\it subdwarf} B (sdB) stars 
discovered by our campaigns of follow-up spectroscopy.
In the Hertzsprung-Russell-Diagram (HRD), sdB stars populate a very narrow area which 
lies on a blueward extension of the horizontal branch (HB), the so called {\it extreme
horizontal branch} (EHB, Heber et al. \cite{Heb84}, Heber \cite{Heb86}, and Saffer et al. 
\cite{Saf94}). They have hydrogen dominated atmospheres 
(ty\-pi\-cal\-ly: $n({\rm He})/n({\rm H})\lesssim 0.01$), with effective temperatures of
$20\,000{\rm K} \le T_{\rm eff} \le 40\,000{\rm K}$ and their logarithmic surface gravities 
are typically between 5.0 and 6.0 (cgs). 
SdB stars consist of a helium-burning core with a canonical mass of $M_{\rm core}\approx 0.5 M_{\sun}$ 
surrounded by a thin hydrogen-rich envelope ($M_{\rm env} < 0.02 M_{\sun}$, Heber \cite{Heb86}, Saffer et al. 
\cite{Saf94}). 
However, their origin is still unclear.

After passing the red-giant stage, these stars must have suffered from such 
a high mass loss rate
that their outer layer was lost almost entirely. The remaining hydrogen-rich envelope has
not enough mass to sustain a hydrogen-burning shell. This means that the star cannot ascend
the asymptotic giant branch (AGB) after the end of the helium-core burning, but should evolve
like a $0.5 M_{\sun}$ helium-main-sequence star (Heber et al. \cite{Heb84}, Heber \cite{Heb86}).
Calculations of Dorman, Rood \& O'Connell (\cite{Dor93}) support this idea.
The reason for very high mass loss at or shortly after the
core helium flash is still unclear and several scenarios are discussed.
As to the origin of sdB stars, a plausible hypothesis is close binary evolution
(Mengel, Norris \& Gross, \cite{Men76}).
In addition to the composite spectrum binaries (Allard et al. \cite{All94}, 
Theissen et al. \cite{The93},\cite{The95} and others) several single-lined binary sdB stars 
have been identified 
from variable Doppler line shifts resulting from orbital motion (Saffer, Livio \&
Yungelson \cite{Saf98}, Maxted, Marsh \& North \cite{Max00}, Maxted et al. \cite{Max01}, and  
Green, Liebert \& Saffer \cite{Gre01}).
At least two thirds of local disk sdB stars are found to be binaries.

SdB stars are important to understand galaxy evolution. 
They are the main cause for the UV excess, the so-called {\it UV upturn}, 
in elliptical galaxies and galaxy bulges (Brown et al. \cite{Bro97}, Brown 
et al. \cite{Bro00b}). The reason is that sdB stars spend a long life time ($\sim 10^8$ years)
on the EHB at high temperatures. 
They are also considered to be useful age indicators for elliptical galaxies
(Brown et al. \cite{Bro00a}). 

\begin{table*}
\caption[]{Observing logs of all HQS follow-up runs for our programme stars.
A range of spectral resolutions is given for the three runs affected by seeing disk 
being temporarily smaller than the slit width.}
\label{observations}
\scriptsize
\centering
\begin{tabular}{rllllll}\hline
run \#  & date                 & instrument & recip. disp. & spectr. res. & wavelength coverage  & observers \\
				& (start of nights)		& 						& [\AA/mm]		 &	[\AA]					& [\AA]									&						\\\hline
 1      & 1989 Jan 21-25       & 3.5m B\&C  & 120        	 &  	5.0						& 3850-5650            & Heber \& Jordan \\
 2      & 1990 Jan 08-17       & 3.5m TWIN  & 144/160     &			6.5						& 3550-5550,5570-7030  & Jordan \& M\"oller  \\
 3      & 1990 Oct 01-09       & 3.5m FR    & 136        	 &		7.0						& 3770-5550            & Heber \& Dreizler \\
 4      & 1990 Nov 04-11       & 3.5m TWIN  & 144/160    	 &		5.5						& 3430-5550,5560-7030  & Jordan \& Rauch \\
 5      & 1991 Jun 19-25       & 3.5m TWIN  & 144/160    	 &		5.5						& 3570-5750,5110-9300  & Heber \& Marten \\
 6      & 1992 Sep 10-14       & 3.5m TWIN  & 144/160     &			4.5-5.5				& 3360-5550,5430-9740  & Dreizler \\
 7      & 1993 Mar 07-12       & 3.5m TWIN  & 144/160    	 &		5.0						& 3470-5680,5420-9630  & Heber \\
 8      & 1993 Aug 28 - Sep 02 & 3.5m TWIN  & 72/72      	 &		3.4						& 3600-5500,5540-7420  & Dreizler \& Haas \\
 9      & 1993 Sep 02-05       & 2.2m CAS   & 120        	 &		4.5-5.5				& 4010-6720            & Haas \& Dreizler \\
10      & 1994 Sep 21-25       & 3.5m TWIN  & 72/72      	 &		3.5						& 3610-5490,5440-7320  & Dreizler \\
11      & 1995 Jan 23-27       & 3.5m TWIN  & 72/72      	 &		3.5						& 3580-5470,5420-7320  & Dreizler \\
12      & 1996 Aug 16-19       & 3.5m TWIN  & 72/72      	 &		3.6						& 3770-5660,5430-7340  & Lemke \\
13      & 1997 Aug 28-31       & 3.5m TWIN  & 72/72      	 &		3.5						& 3300-5450,5300-7550  & Edelmann \\
14      & 1998 Sep 30 - Oct 04 & 2.2m CAFOS & 100        	 &		5.0-8.0				& 3400-6300            & Edelmann \\\hline
\end{tabular}
\end{table*}
Subdwarf B stars are also very important in the context of stellar astrophysics.
The discovery of several pulsating sdB stars (called sdBV or EC 14026 stars, after 
the prototype EC\,14026$-$2647, Kilkenny et al. \cite{Kil97}) has rapidly increased the 
interest in these objects, because of the prospect of probing their structure 
by asteroseismo\-logy. 
The driving mechanism of the pulsation is due to an opa\-ci\-ty bump associated with 
an iron ionization in the envelopes of sdB stars (Charpinet et al. \cite{Cha96}, \cite{Cha97}).
The prediction of Charpinet et al. (\cite{Cha97}) that sdB stars in the temperature range of 
$29\,000{\rm K} \le T_{\rm eff} \le 36\,000{\rm K}$ should pulsate is very well
confirmed by subsequent spectros\-co\-pi\-cal analyses of the EC 14026 stars 
(Heber, Reid \& Werner \cite{Heb00}, {\O}stensen et al. \cite{Ost01a}, \cite{Ost01b}, 
Dreizler et al. {\cite{Dre02}, Silvotti et al. \cite{Sil02}).
Now, 29 pulsating sdB stars are known (see O'Donoghue et al. \cite{ODo99} 
and Charpinet \cite{Cha01} for reviews).

For all these investigations, knowledge of the stellar parameters is very important.
We present here the results of a spectral analysis of a large sample of sub\-dwarf B stars 
selected from follow-up observations of candidates from the Hamburg Quasar Survey.
\section{Programme stars}
\subsection{Preselection}
Candidate stars were selected from the HQS objective prism plates, 
first by automatically selecting spectra  which are blue compared to
the bulk of spectra, and second, by visually classifying them on a
graphics display (Hagen et al. \cite{Hag99}). Objects with  UV excess emission
were either classified directly as hot stars due to the presence
of hydrogen Balmer absorption lines or were retained as quasar candidates.
Follow-up observations of stellar candidates were mostly made from the
hot stars list, and occasionally were of objects that were  not confirmed as 
quasars. 
\subsection{Observations}
The observations described here were obtained, starting in January 1989, at the DSAZ on Calar Alto, 
Spain, during various observing runs with different telescopes and instruments.
The observational dataset is therefore very inhomogeneous. The spectral resolution ranges from 
3.4\AA\ to 8.0\AA\ FWHM (determined from the FWHM of the helium-argon lines in the calibration
spectra) and the wavelength coverage also varies between 4010--6720\AA\ and 3360--9740\AA.
For three observing runs the seeing was sometimes better (smaller) than the slit width, resulting
additionally in a varying spectral resolution during these runs.
For a detailed overview of the observational material see Table 
\ref{observations}.
\subsection{Data reduction}
The spectra were extracted from the two-dimensional frames and reduced to linear wavelength and 
intensity scales using the IDAS package written by G. Jonas in Kiel for the early observations
(until 1991) and the ESO-MIDAS package for the data obtained after 1991.

All frames were bias subtracted, flat field corrected, and cosmic ray events were cleaned.
The sky background was removed by extracting a stripe on each side of 
the star's spectrum and subtracting the
average of these two stripes from each row of the stellar signal on the CCD.
These corrected rows were combined to a one dimensional stellar spectrum.
Thereafter a wavelength calibration was performed 
with calibration spectra recorded immediately after each stellar spectrum.
Then all wavelength-calibrated spectra were corrected for atmospheric extinction using the 
extinction coefficients of La Silla, Chile  
(T\"ug \cite{Tug77}) as these coefficients are not available for the Calar Alto 
observatory.
In the last step all spectra were relatively flux calibrated using spectra of flux-standard 
stars (mostly BD+28\degr\,4211, G\,191$-$B2B or Feige\,34, Oke \cite{Oke90}) 
which were taken each night.

A subset of spectra obtained is presented in Fig. \ref{samplespectra}.
The object list is supplemented by 
one spectrum of a sdB star (\object{HS\,1641+4601}) kindly provided by T. Rauch.
\subsection{Selection and classification}
From the list of stellar HQS follow-up observations,
we selected here 111 subdwarf B candidates for a detailed analysis, 
using the classification system of Moehler et al. (\cite{Moe90b}):
 The optical spectra of subdwarf B stars are dominated by strong broad Balmer lines of 
neutral hydrogen and weak or absent He {\sc i} lines. 
The so-called sdOB stars, introduced by Baschek \& Norris (\cite{Bas75}), represent a hotter 
group of the sdB stars, that show in addition to strong broad Balmer and weak He~{\sc i} lines 
a weak He~{\sc ii} 4686\AA\ absorption line in their spectra.

A closer inspection revealed 18 sdB stars of the sample
to be spectroscopic binaries. All of them show at least two 
characteristics of a cool companion star
(e.g. flat flux distribution, G-band absorption, Ca H+K, Mg {\sc i} triplet at 5167\AA, 
5173\AA, and 5184\AA) (see Table \ref{bintab}).
The spectral classifications for all programme stars  are listed in Table~4.

The coordinates were determined on
HQS direct plates and are accurate to $\pm$2\arcsec. We checked  the Digital Sky Survey for
all stars and found that usually the object cannot
be mistaken. Stars nearby were found in eight cases and for those we
present finding charts in Fig. \ref{findingcharts}.
The B-magnitudes presented in Table \ref{results} were determined mostly 
from the objective prism plates and may have an error of up to 0.3 mag
except when marked by a colon ($\pm 0.5$ mag uncertainty).

Several stars have already been discovered as UV excess objects independently by various
surveys.
The re\-fe\-ren\-ces are indicated in Table \ref{results}.
\begin{table}
\caption[]{Spectral signatures of cool companion stars in the spectroscopic binaries
  of our sample.}
\scriptsize
\centering
\label{bintab}
\begin{tabular}{lcccc} \hline
binary stars & Ca H+K & G-band & Mg {\sc i} & flat flux \\ \hline
\object{HS 0028+4407} &   $\sqrt{}$    &   $\sqrt{}$    &    $\sqrt{}$      &  $\sqrt{}$  \\
\object{HS 0127+3146} &   -            &   $\sqrt{}$    &    $\sqrt{}$      &  -  \\
\object{HS 0136+0605} &   -            &   -            &    $\sqrt{}$      &  $\sqrt{}$  \\
\object{HS 0215+0852} &   -            &   $\sqrt{}$    &    $\sqrt{}$      &  $\sqrt{}$  \\
\object{HS 0252+1025} &   -            &   $\sqrt{}$    &    $\sqrt{}$      &  $\sqrt{}$  \\
\object{HS 0446+1344} &   -            &   $\sqrt{}$    &    $\sqrt{}$      &  $\sqrt{}$  \\
\object{HS 0656+6117} &   -            &   $\sqrt{}$    &    $\sqrt{}$      &  -  \\
\object{HS 0942+4608} &   $\sqrt{}$    &   $\sqrt{}$    &    $\sqrt{}$      &  $\sqrt{}$  \\
\object{HS 1106+6051} &   $\sqrt{}$    &   $\sqrt{}$    &    $\sqrt{}$      &  $\sqrt{}$  \\
\object{HS 1511+6221} &   $\sqrt{}$    &   $\sqrt{}$    &    $\sqrt{}$      &  -  \\
\object{HS 1612+6337} &   $\sqrt{}$    &   $\sqrt{}$    &    $\sqrt{}$      &  $\sqrt{}$  \\
\object{HS 1612+7335} &   $\sqrt{}$    &   $\sqrt{}$    &    $\sqrt{}$      &  $\sqrt{}$  \\
\object{HS 1615+6341} &   -            &   $\sqrt{}$   &    $\sqrt{}$      &  $\sqrt{}$  \\
\object{HS 1753+5342} &   $\sqrt{}$    &   $\sqrt{}$    &    $\sqrt{}$      &  $\sqrt{}$  \\
\object{HS 1753+7025} &   $\sqrt{}$    &   $\sqrt{}$    &    $\sqrt{}$      &  $\sqrt{}$  \\
\object{HS 1844+5048} &   -            &   $\sqrt{}$    &    $\sqrt{}$      &  -  \\
\object{HS 1858+5736} &   $\sqrt{}$    &   -            &    $\sqrt{}$      &  $\sqrt{}$  \\
\object{HS 2216+1833} &   $\sqrt{}$    &   $\sqrt{}$    &    $\sqrt{}$      &  $\sqrt{}$  \\ \hline
\end{tabular}
\end{table}
\normalsize 
Although 45 of our programme stars can be found in other surveys, we found 
spectral classifications for only 21 of them in the literature.
As can be seen in Table \ref{class} there is a good agreement with previous
classifications except for \object{HS\,2333+3927} which was previously classified 
as a DAZ white dwarf.
\begin{table}
\caption[h]{Comparison of spectral classifications with previous work.
	Since the PG classification scheme differs from ours, we transcribed the
  PG spectral types into our scheme (Moehler et al. \cite{Moe90b}).} 
\label{class}
\scriptsize
\centering
\begin{tabular}{cccl}\hline
star          &  this work  & other 		& Ref. \& name within\\\hline
\object{HS~0016+0044}  &  sdB        & sdB   		& T94 \\
\object{HS~0039+4302}  &  sdB        & sdB   		& B91: \object{Balloon~84041013}\\
\object{HS~0048+0026}  &  sdOB       & sdB   		& B91: \object{Balloon~94700002}\\
              &             & sdOB  		& G86: \object{PG~0048+004}\\ 
\object{HS~0055+0138}  &  sdB        & sdB   		& B91: \object{Balloon~94700001}\\
							&							& sdB   		& P86: \object{PG~0055+016}\\ 
\object{HS~0209+0141}  &  sdB        & sdB   		& G86: \object{PG~0209+0141}\\
\object{HS~0212+1446}  &  sdB 				& sdB				& G86: \object{PG~0212+148}\\
\object{HS~0232+3155}  &  sdOB       & sdB  		 	& W90: \object{KUV~02324+3156}\\
\object{HS~0941+4649}  &  sdB				& sdB   		& M98: \object{US~909}\\
\object{HS~0942+4608}	&  binary			& sdB+G 		& H89\\
							&							& sdB				& M98: \object{PG~0942+461}\\
\object{HS~1106+6051}	&	 binary			& sdB				& G86: \object{PG~1106+608}\\
\object{HS~1236+4745}	&  sdB				& sdB				& S94: \object{PG~1236+479}\\
\object{HS~1511+6221}  &  binary     & sdB+K5		& A94: \object{PG~1511+624}\\
							&							& sdB				& A96: \object{FBS~1511+624}\\
\object{HS~1547+6312}  &  sdB				& sdB				& A96: \object{FBS~1547+632}\\
							&							&	sdB				& G84: \object{PG~1547+632}\\
\object{HS~1612+7335}	&  binary			& sdB+K2.5	& A94: \object{PG~1612+735}\\
\object{HS~1641+4601}  &  sdB        & sdB   		& B91: \object{Balloon~83600002}\\
\object{HS~2218+0201}	&	 sdB				& sdB				& U98: \object{PG~2218+020}\\
\object{HS~2233+2332}	&	 sdOB				& sdB				& B91: \object{Balloon~90900003}\\
\object{HS~2240+0136}  &  sdB				& sdOB?			& K84: \object{PHL~384}\\
\object{HS~2240+1031}	&  sdOB				& sdB				& G84: \object{PG~2240+105}\\
\object{HS~2246+0158}	&  sdB				& sdB				& G84: \object{PG~2246+019}\\
\object{HS~2333+3927}  &  sdOB       & DAZ   		& A96: \object{FBS~2333+395}\\\hline
\end{tabular}\\[1mm]
\small
\parbox[t]{85mm}{Ref.:
		A94 = Allard et al. (\cite{All94});
    A96 = Abrahamian \& Mickaelian (\cite{Abr96});
    B91 = Bixler, Bowyer \& Laget  (\cite{Bix91});
		G86 = Green, Schmidt \& Liebert (\cite{Gre86});
		H89 = Heber, Jordan \& Weidemann (\cite{HJW89});
		K84 = Kilkenny (\cite{Kil84});
		M98 = Mitchell (\cite{Mit98});
		R93 = Rodgers, Roberts \& Walker (\cite{Rod93});
		S94 = Saffer et al. (\cite{Saf94});
		T94 = Thejll, Theissen \& Jimenez (\cite{Thej94});
		W90 = Wegner, Steven \& Swanson (\cite{Weg90}).}
\normalsize
\end{table}
\section{Atmospheric parameters}
\begin{figure*}
\vspace{11.5cm}
\includegraphics{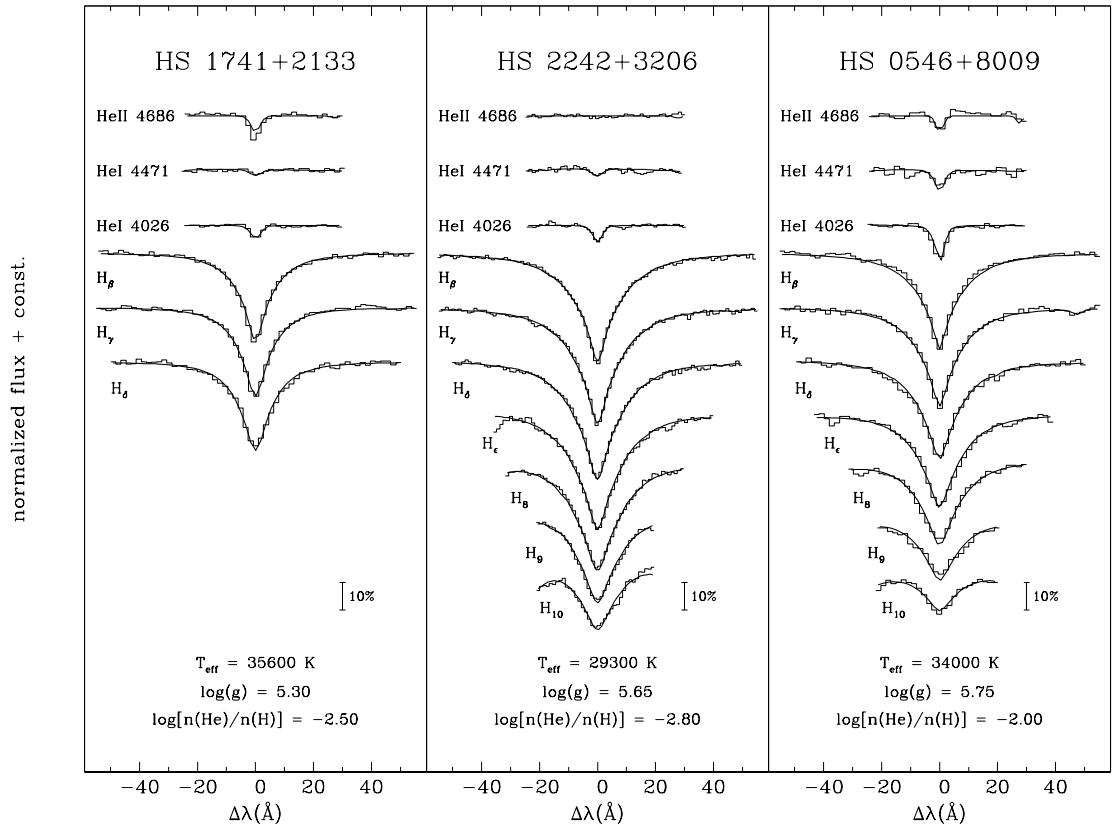}
	\caption[]{Sample fits for three programme sdB stars. The observed spectra are plotted as
			histograms. A detailed discussion of the fit for HS~1741+213 is given in section 3.2. 
 	\label{fit}
}
\end{figure*}
The stellar atmospheric parameters (effective temperature, surface gravity, and 
photospheric helium abundance) were determined by matching synthetic line profiles 
calculated from model 
atmospheres to all Balmer (mainly ${\rm H}_{\beta}$ up to ${\rm H}_{\epsilon}$) and helium 
(mainly \ion{He}{i} $\lambda\lambda$ 4026\AA, 4471\AA, 5015\AA, 5876\AA, and \ion{He}{ii} 4686\AA) 
line profiles present in the observed spectra of all programme stars.
\subsection{Model atmospheres and synthetic spectra}
Three different sets of models were used:
\begin{enumerate}
	\item	A grid of metal-line blanketed LTE model atmospheres (Heber et al. \cite{Heb00}).
			The models are plane parallel and chemically homogeneous
			and consist of hydrogen, helium, and metals (solar abundances).
			The synthetic spectra were calculated with LINFOR\footnote{LINFOR was originally
      developed by H.~Holweger, M.~Steffen, and W.~Steenbock at Kiel University. It has been
      enhanced and maintained by M.~Lemke, with additional mo\-di\-fi\-cations by N.~Przybilla.
      For a description see:\\ 
      {\tt \scriptsize http://www.sternwarte.uni-erlangen.de/$\sim$ai26/linfit/linfor.html}}. 

			For the spectrum synthesis, line profiles were calculated for the
			Balmer series of neutral hydrogen ($n$ up to 22) with Stark broadening tables of 
			Lemke (\cite{Lem97b}) which uses the unified theory of Vidal et al. (\cite{Vid73}). 
			Helium lines were calculated using broadening tables of Barnard, 
			Cooper \& Smith (\cite{BCS74}), Shamey (\cite{Sha69}), and 
      Griem (\cite{Gri74}) for He {\sc i}, and 
			Sch\"oning \& Butler (\cite{Sch89}) for He {\sc ii}.
			The metal line blanketing was included by the opacity distribution function (ATLAS6) of 
			Kurucz (\cite{Kur79}).
			The grid covers the area for EHB Stars:
			$T_{\rm eff}= 11\,000{\rm K} \ldots 35\,000$K 
			in steps of $\Delta T_{\rm eff} = 1\,000$K to $2\,500$K; 
			$\log(g) = 3.50 \ldots 6.50$ [{\it cgs}] in steps of $\Delta\log(g)=0.25$;
			$n({\rm He})/n({\rm H})=0.0001, 0.001, 0.01, 0.03, 0.10, 0.33$. 
	\item A grid of partially line blanketed NLTE model atmospheres (Napiwotzki \cite{Nap97}).
			The models are plane parallel, chemically homogeneous, and consist of
			hydrogen and helium. The latest version of the NLTE code from Werner (\cite{Wer86})
			was used which is based on the Accelareted Lambda Iteration (ALI) of
			Werner \& Husfeld (\cite{WHu85}).	
			The following grid was used: $T_{\rm eff}=(27, 30, 32, 35, 37, 40, 45)\times 1000$K;
			$\log(g) = 3.50 \ldots 6.50$ in steps of $\Delta\log(g)=0.25$;
			$n({\rm He})/n({\rm H})= 3\cdot 10^{-4}, 1\cdot 10^{-3}, 3\cdot 10^{-3}, 0.01, 0.03, 0.1, 0.3$.
	\item A grid of partially line blanketed NLTE model atmospheres 
      for He-rich ($n({\rm He})/n({\rm H}) > 0.3$) objects. 
			An extended and updated grid of Dreizler et al. (\cite{Dre90}), based on 
      the ALI method,  
      was used.
			The models are plane parallel and chemically homogeneous and consist of
			hydrogen and helium. 
			The grid covers: $T_{\rm eff}=(35, 40, 45)\times 1000$K;
			$\log(g) = 4.0 \ldots 6.5$ in steps of $\Delta\log(g)=0.5$;
			$n({\rm He})/n({\rm H})= 0.5, 1, 3, 10, 100$.
\end{enumerate}
\subsection{Fit procedure}
The matching of the observed spectra was done by means of a $\chi^2$ fit
using an updated procedure of Bergeron et al. (\cite{Ber92}) and Saffer et al. (\cite{Saf94})
which determines simultaneously the atmospheric parameters.
Beforehand all spectra were normalized and 
the model spectra were folded with the instrumental profile (Gaussian with appropriate 
width).
Rotational broadening was neglected in the fitting procedure.
Some fit examples are shown in Fig. \ref{fit}.
			
The numbers of Balmer lines that can be used for the analysis may be limited by insufficient
spectral coverage. Hence several stars are left with only three Balmer lines (see Fig. \ref{fit}).
In order to check whether the results depend on the number of Balmer lines included in the fit, we
compared results based on many Balmer lines to those from three lines for stars with sufficient 
spectral coverage. No systematic differences became apparent.

The fit reproduces the Balmer lines well. For the hottest stars the \ion{He}{i}/\ion{He}{ii}
ionisation equilibrium provides an additional temperature indicator.
In most cases (e.g. HS~0546+8009, see Fig. \ref{fit}) the fit of the \ion{He}{i} and \ion{He}{ii} lines 
is consistent with that of the Balmer lines.
However, for sdOB stars showing a \ion{He}{ii} 4686\AA\ line which is comparable or stronger than
the \ion{He}{i} 4471\AA\ line (i.e. for HS~0048+0026, HS~1051+2933, HS~1741+2133, HS~2156+2215, and 
HS~2333+3927),
the helium ionisation equilibrium indicates a considerably higher effective temperature than 
from the Balmer lines.
The most extreme case is HS~1741+2133 displayed in Fig. \ref{fit} (left panel).
To match the \ion{He}{ii} line an effective temperatute larger by $\approx 3000$K would be required.
Such a discrepancy has also been observed in the analysis of high resolution spectra of the
pulsating sdB star PG~1219+534 (Heber, Reid \& Werner \cite{Heb00}). 
A detailed discussion is given in that paper.
In the absence of an explanation for this helium line problem we adopt the parameters from the 
fit of all lines (H+He).

Our fit process, however, fails in the case of composite spectra.
Without knowledge of the flux distributions of the cool companions it is 
impossible to extract the spectra of the sdB stars. To analyse these binaries, 
additional spectra and more sophisticated 
procedures are necessary (see e.g. Aznar Cuadrado \& Jeffery \cite{Azn02}).
We defer the analysis of the composites to a subsequent paper.
\subsection{Results}
\begin{figure}
\vspace{8.3cm}
\includegraphics{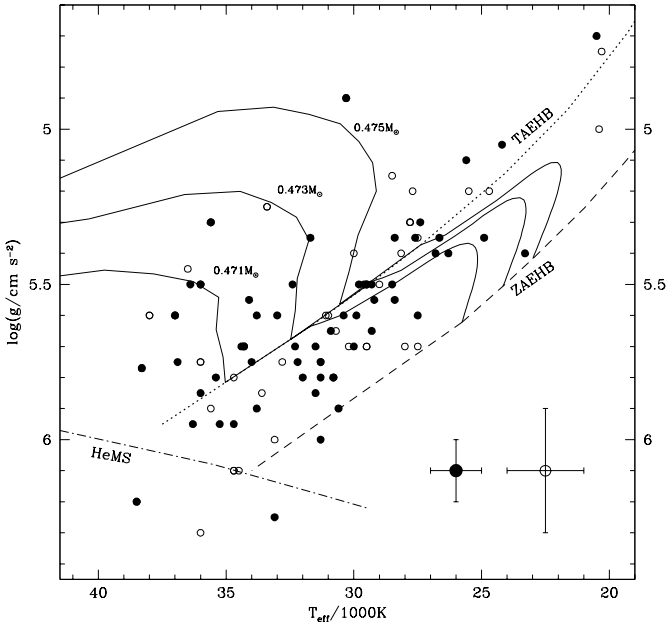}
\caption[]{Distribution of our sdB sample in the $T_{\rm eff}$--$\log(g)$--plane.
Stars denoted by open circles have larger uncertainties for the derived temperatures and gravities 
than the stars denoted by filled circles (see typical error-bars at the lower right side)
mostly due to lower S/N and/or spectral resolution.
The Helium Main Sequence (HeMS, Paczy\'nski \cite{Pac72}) together with
the Zero Age Extended Horizontal Branch for solar metallicity (ZAEHB), the Terminal Age 
Extended Horizontal Branch for solar metallicity (TAEHB), and evolutionary tracks
for extended horizontal branch stars from Dorman, Rood \& O'Connell (\cite{Dor93})  
for three different masses (solar metallicity) are shown for comparison.
\label{hrd}
}
\end{figure}
\begin{figure}
\vspace{8.3cm}
\includegraphics{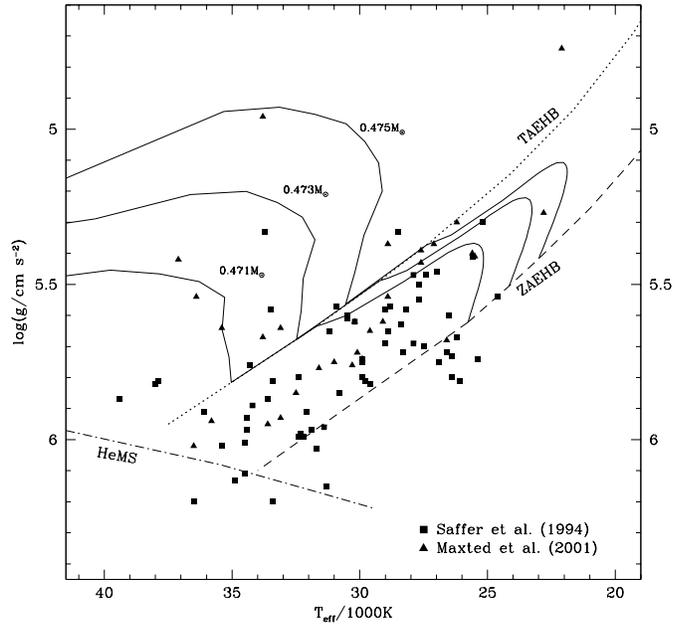}
\caption[]{Distribution of the sdB samples of Saffer et al. (\cite{Saf94} filled squares)
and Maxted et al. (\cite{Max01}, filled triangles) in the $T_{\rm eff}$--$\log(g)$--plane
for comparison purposes (see sections 3.4 and 5).
\label{hrd_saf_max}
}
\end{figure}
Table \ref{results} summarizes the results 
of our analysis 
including spectral types, effective temperatures, gravities, and helium abundances. 
Additionally, the equatorial and galactical coordinates, 
the $B$ magnitudes and extinctions, the radial velocities (see section 4.1), the absolute visual
magnitudes, the distances from earth and from the galactic plane (see section 4.2), and
the references are given for all programme stars.
The values given with $27\,000\mbox{K} \le T_{\rm eff}\le 35\,000$K 
represent mean results of our LTE and non-LTE fits.
All values with  $T_{\rm eff}< 27\,000$K are exclusively from LTE, and all values with
$T_{\rm eff}> 35\,000$K are solely from non-LTE fits.
Statistical errors for the atmospheres which are derived from the fit program
are unrealistically small
(typically: $\sigma_{T_{\rm eff}}^{\rm fit}\approx 300$K, $\sigma_{\log(g)}^{\rm fit}\approx 0.05$dex,
$\sigma_{\log[n({\rm He})/n({\rm H})]}^{\rm fit}\approx 0.05$dex).
The syste\-ma\-ti\-c errors that arise from the observations (spectral resolution, S/N), and 
from the data reduction (flat-field correction, background subtraction, relative flux calibration, 
and continuum placement) 
are dominant. The real errors can only be estimated.
Individual error estimates for the effective temperatures, the gravities, and helium abundances 
are given in Table \ref{results}.
Four EHB programme stars which are observed twice at different dates with different
instruments allow a selfconsistency check. 
As can be seen, the results match well within the given error limits.
In these cases the mean results are plotted in Fig. \ref{hrd}.

The analysis shows that 89 ($\sim$96\%) of the 93 selected apparently single stars are bona fide sdB or 
sdOB stars.
One (\object{HS\,2229+0910}) is considered to be a blue horizontal branch (HBB) star,
while three stars (\object{HS\,0231+8019}, \object{HS\,1556+6032}, and \object{HS\,2131+0349}) 
have atmospheric parameters consistent with those of normal main sequence B stars.
One of the sdOB stars (\object{HS\,1051+2933}) is identified as unusually 
helium rich, i.e. exceeding the solar helium abundance.
The results for all apparently single programme sdB stars 
are also shown in Fig. \ref{hrd}  in a $T_{\rm eff}$--$\log(g)$--diagram.
For comparison we plot the results of the analyses of sdB stars by Saffer et al. (\cite{Saf94}) and 
Maxted et al. (\cite{Max01}) in Fig. \ref{hrd_saf_max}. The further discussion is deferred to 
section 5.
\subsection{Correlations of the helium abundance with stellar parameters?}
In order to search for possible trends of the chemical composition with the 
atmospheric parameters 
we compare the derived helium abundances with the measured stellar 
parameters $T_{\rm eff}$, $\log(g)$,
and with the luminosity\footnote{We express the luminosity in terms of the 
Eddington Luminosity $L_e$ (for electron scattering, see Eq. \ref{luminosity_f}).}
for all apparently single EHB programme stars in Figs. \ref{nhe_teff},  
\ref{nhe_logg}, and \ref{logl_le__loghe_h}.
\begin{figure}
\vspace{7.2cm}
\includegraphics{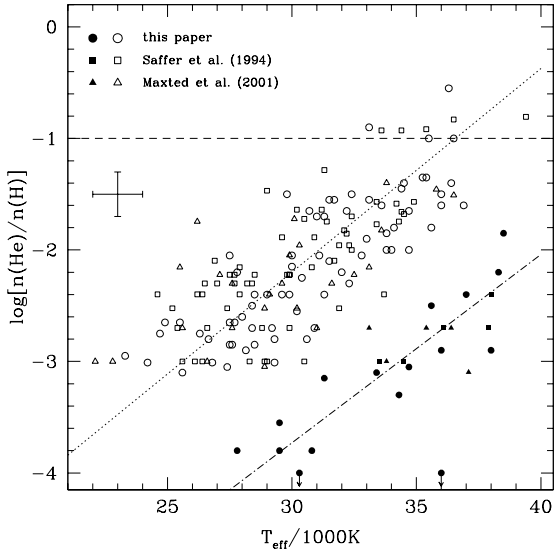}
\begin{minipage}{8.8cm}
	\caption[]{Plot of the helium abundance versus effective tem\-pe\-ra\-tu\-re. 
  Additionally the results of Saffer et al. (\cite{Saf94}, squares) 
  and Maxted et al. (\cite{Max01}, triangles) are plotted.
  The dotted line indicates the linear regression (Eq. \ref{reg1}) for the bulk 
  of the sdB stars (open symbols)
  and the dashed-dotted line shows the linear regression (Eq. \ref{reg2})
  for the peculiar sdB stars
  (filled symbols). 
  The dashed horizontal line denotes the solar helium abundance. 
 	\label{nhe_teff}
}
\end{minipage}
\end{figure}
\begin{figure}
\vspace{7.2cm}
\includegraphics{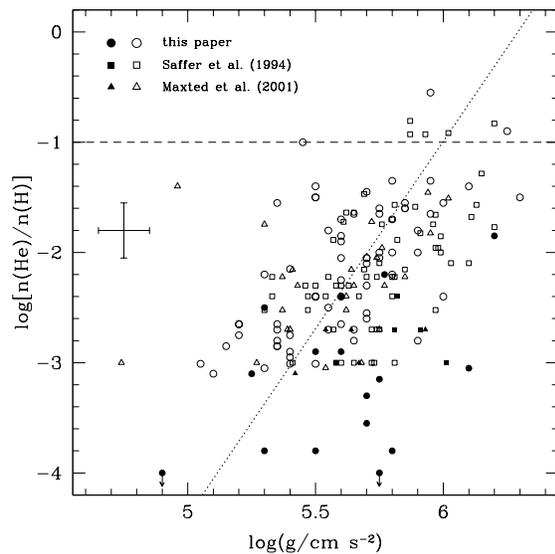}
\begin{minipage}{8.8cm}
	\caption[]{Plot of the helium abundance versus gravity.
  Additionally the results of Saffer et al. (\cite{Saf94}, squares) 
  and Maxted et al. (\cite{Max01}, triangles) are plotted.
   The dashed horizontal line denotes the solar helium abundance.
  For the filled symbols cf. Fig. \ref{nhe_teff}.
  The dotted line is the linear regression for the bulk of the sdB stars 
 (open symbols).
 \label{nhe_logg}
}
\end{minipage}
\end{figure}
\begin{figure}
\vspace{7.2cm}
\includegraphics{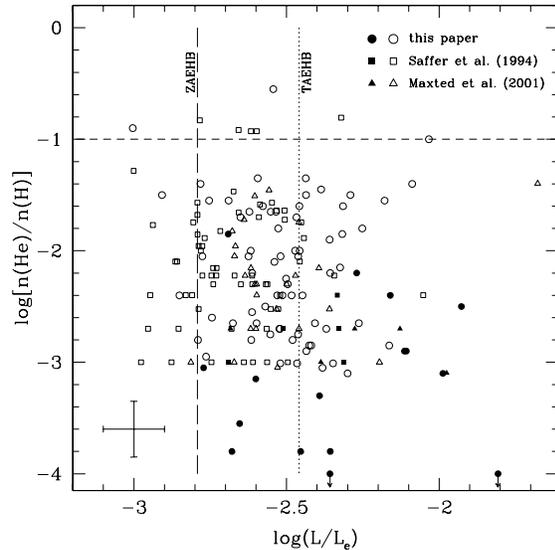}
\begin{minipage}{8.8cm}
	\caption[]{Plot of the helium abundance versus luminosity (with respect to the Eddington
   luminosity $L_e$, see Eq. \ref{luminosity_f}).
  Additionally the results of Saffer et al. (\cite{Saf94}, squares) 
  and Maxted et al. (\cite{Max01}, triangles) are plotted.
   The dashed horizontal line denotes the solar helium abundance,
   the long dashed vertical line denotes the ZAEHB, and the dotted vertical line
   the TAEHB.
  For the filled symbols cf. Fig. \ref{nhe_teff}.
 	\label{logl_le__loghe_h}
}
\end{minipage}
\end{figure}

First of all, we discovered a clear correlation between the helium abundance
and the effective temperature ($T_{\rm eff}$):
The larger the temperature, the larger the helium content (cf. Fig. \ref{nhe_teff}).
Furthermore, there seems to be a separation into two sequences of sdB stars
in the $T_{\rm eff}$ -- helium abundance plane.
A fraction of our ana\-lysed sdB stars (about 1/6th, indicated with filled symbols)   
have much lower helium abundances at the same temperatures than the bulk of the sdB
stars.

Fig. \ref{nhe_logg} may indicate a connection 
between the helium abundance and the gravity.
However, for sdB stars the gravity is not independent of the effective temperature 
(see Eq. \ref{luminosity_f}) since the horizontal branch is a sequence of nearly
constant luminosity.
The stars that seperate from the main bulk (filled circles) 
in Fig. \ref{nhe_teff} lie somewhat below the main bulk in Fig. \ref{nhe_logg} as well,
but do not separate as clearly as in the former diagram.

The luminosity as derived from gravity and  $T_{\rm eff}$
\begin{eqnarray}
L/L_e = T_{\rm eff}^4/(10^{15.118}\times g) \label{luminosity_f}
\end{eqnarray}
is plotted in Fig. \ref{logl_le__loghe_h}.
No correlation is detectable.
However, for the peculiar sdB stars (indicated with filled symbols) there is a
slight tendency for a correlation to occur at higher luminosities.

To verify our discoveries, we have searched in the literature for other
analyses which determined the atmospheric parameters 
using a method similar to ours.
The results of
Saffer et al. (\cite{Saf94}, squares) who analysed 68 EHB stars 
and those of Maxted et al. (\cite{Max01}, triangles) who analysed
36 EHB stars for atmospheric parameters
are added to ours 
in Figs. \ref{nhe_teff} to \ref{logl_le__loghe_h}.
The correlation between the helium abundance and the effective temperature is confirmed. 
Furthermore, 
the suggested separation into two sequences (cf. Fig. \ref{nhe_teff})
is reinforced.

A linear regression for the bulk  of sdB stars (open symbols) gives:
\begin{equation}
 \log\left[\frac{n({\rm He})}{n({\rm H})}\right] = -3.53 + 1.35 \cdot \left( \frac{T_{\rm eff}}{10^4\mbox{K}} - 2.00 \right) 
\label{reg1}  
\end{equation}
For the other sequence (filled symbols, except the two upper limit values indicated by 
downward arrows) we get:
\begin{equation}
 \log\left[\frac{n({\rm He})}{n({\rm H})}\right] = -4.79 + 1.26 \cdot \left( \frac{T_{\rm eff}}{10^4\mbox{K}} - 2.00 \right)  
\label{reg2}  
\end{equation}
\subsection{Comparison with previous results}
Spectroscopic analyses are available in the literature only for five of our programme 
stars from three different groups (Moehler et al. \cite{Moe90a}, Bixler, Bowyer \& Laget \cite{Bix91},
and  Saffer et al. \cite{Saf94}). 

Different methods were applied to determine the stellar parameters:
Two groups (Bixler, Bowyer \& Laget \cite{Bix91} and  Saffer et al. \cite{Saf94}) 
used a procedure similar to the one described here
(fitting of model line profiles to optical spectra)
to derive $T_{\rm eff}$ and $\log(g)$. 
Saffer et al. included the determination of the helium abundance 
into their fit process,
whereas Bixler et al. derived the helium abundance from equivalent width
measurement of the \ion{He}{i} 4471\AA, 4922\AA, and \ion{He}{ii} 4686\AA\ lines.
Moehler et al. (\cite{Moe90a}) used a three-step-procedure:
The effective temperature has been calculated first from 
colour indices.
Keeping the temperature fixed, the surface gravity was obtained
by visual comparison of model line profiles with optical spectra
of one or more Balmer lines (mainly H$_\gamma$). 
Finally, the helium abundance was derived by measuring the equivalent width 
of the \ion{He}{i} 4471\AA\ line.

The sample of Bixler, Bowyer \& Laget (\cite{Bix91}) overlaps with ours for three stars
(\object{HS~0039+4302}, \object{HS~1641+4601}, and \object{HS~2233+2332}).
However, the results given in Bixler, Bowyer \& Laget (\cite{Bix91}) suffer from
very large error limits ($\Delta T_{\rm eff}/T_{\rm eff}\approx 15-20\%
$,
$ \Delta\log(g)\approx 0.4 - 0.7$dex) probably due to the low resolution and S/N 
of their spectra,
which renders a comparison with our results useless.

One star (\object{HS~0212+1446}) overlaps with the sample of Moehler et al. (\cite{Moe90a}).
The values differ considerably: the effective temperature determined by Moehler et al. (\cite{Moe90a})
is 5\,000K lower and the gravity is 0.9dex lower than our results. 
Saffer et al. (\cite{Saf94}),
who discovered similar differences comparing their results with that of 
Moehler et al. (\cite{Moe90a}), 
argue that the calibration of the Str\"omgren colours 
used by Moehler et al. (\cite{Moe90a}) is inappropriate for sdB stars 
and causes larger systematic errors.
This view is supported by investigations of Napiwotzki, Sch\"onberner \& Wenske (\cite{Nap93}).

There remains only one sdB star of our sample that
can be compared with the results of another group.
Saffer et al. (\cite{Saf94}) determined the stellar parameters for \object{HS~1236+4754}
to be $T_{\rm eff}=27\,900\mbox{K}\pm1\,000\mbox{K}, \log(g)=5.47\pm 0.15$, and 
$n({\rm He})/n({\rm H})=0.004$, which is in perfect agreement with our 
result: $T_{\rm eff}=28\,400\mbox{K}\pm800\mbox{K}, \log(g)=5.55\pm 0.10$, and 
$n({\rm He})/n({\rm H})=0.003$.
\section{Radial velocities and distances}
\subsection{Radial velocities}
In view of the large fraction of single lined binaries among the sdB stars 
(Maxted et al. \cite{Max01}) it is worthwhile to measure radial velocities (RVs) of our 
programme stars.
The RVs are determined by calculating the shifts of the measured wavelengths 
of all fitted Balmer and helium lines relative to laboratory wavelengths.
Afterwards they are corrected to heliocentric values.
We decided to determine the RVs only for spectra with 
spectral resolutions equal to or better than 3.6\AA, because the error margins for the 
spectra of lower re\-so\-lu\-tion are too large to yield meaningful results.

The resulting values are accurate to about $\pm 30$km/s and can be found in Table \ref{results}.
Out of four stars which were observed twice at different dates, only one (\object{HS\,2333+3927}) 
is found to be a RV variable.
The resulting velocities for the other sdB stars, which are observed only once, are
given in Table \ref{results} for comparison with future RV measurements.
\subsection{Distances}
We calculated the distances for all sdB programme stars,
assuming a canonical mass of 0.5 $M_{\sun}$.
From the derived gravities, the radii of the stars are calculated.
By comparing the model atmosphere flux with the dereddened 
apparent visual magnitude\footnote{
The visual magnitudes are estimated from the $B$ magnitudes by adding the typical intrinsic
colour of {$B-V=-0.28$}~mag (Altmann, Edelmann \& deBoer, in prep.).
}, the angular diameter of a star could be determined.
Because several of the programme stars lie at relatively low galactic latitudes 
(20\degr$<|b_{\rm II}|<$30\degr), interstellar reddening can be significant, e.g.
as high as $E(B-V)=0\fm3$ for \object{HS\,0357+0133} (see Table \ref{results}).
All stars lie at distances between 300pc and several kiloparsec from the galactic plane,
therefore beyond the galactic dust layer.
The reddening for all programme stars are estimated from the maps of Schlegel et al. (\cite{Schl98}).
From radii and angular diameters, the distances of the stars follow immediately.
The distance $z$ from the galactic plane is derived from
the galactic latitudes: $z = d \cdot \sin(b_{\rm II})$.     
The results are listed in Table \ref{results}.

Three programme stars are apparently normal B stars. Assuming that they are main sequence stars, 
we derive masses for \object{HS~0231+8019}, \object{HS~1556+6032}, and \object{HS~2131+0349} 
of 5.5$M_{\sun}$, 4.2$M_{\sun}$, and 5.0$M_{\sun}$, respectively, 
using the procedure of Ramspeck, Heber \& Moehler (\cite{Ram01a}).
Using these masses we get distances $d$ (and $|z|$) of $\approx 21$kpc (7kpc) for HS~0231+8019,
$\approx 21$kpc (11kpc) for HS~2131+0349, and $\approx 27$kpc (19kpc) for HS~1556+6032.
The z distances determined for HS~0231+8019 and HS~2131+0349 are not extraordinary in comparison 
to other known apparently normal B stars at high latitudes which are closer than about 10kpc from
the galactic plane (Rolleston et al. \cite{Rol99}, Ramspeck, Heber \& Moehler \cite{Ram01a}).
However, HS~1556+6032 seems to be clearly farther away than  other known apparently
normal B stars in the halo of our Milky Way.
\section{Discussion}
In the gravity versus effective temperature plane (Fig. \ref{hrd}), our confirmed sdB and sdOB
stars lie in a region close to the EHB, but with a tendency to cluster near the
TAEHB when the He-core burning diminishes and the phase of helium shell burning starts.
However, according to the evolutionary life times, most stars should be found 
closer to the ZAEHB,  
like seen in Heber (\cite{Heb86}, Fig. 6),
Saffer et al. (\cite{Saf94}, Fig. 5), and Maxted et al. (\cite{Max01}, Fig. 2).
Comparing our results (Fig. \ref{hrd}) to that of Saffer et al. (\cite{Saf94}) and/or 
Maxted et al. (\cite{Max01}) (see Fig. \ref{hrd_saf_max}), a systematic difference can be suggested.
Saffer et al. (\cite{Saf94}) even found some sdB 
stars (mostly at low temperature, $T_{\rm eff}=25\,000$ to $27\,000$K) to lie below
the ZAEHB (cf. Fig. \ref{hrd_saf_max}).
Because only one star is common to in both studies (both sets of parameters agreed very well, 
see section 3.4), a direct comparison was possible for
this case only.
\begin{figure}
\vspace{7.2cm}
\includegraphics{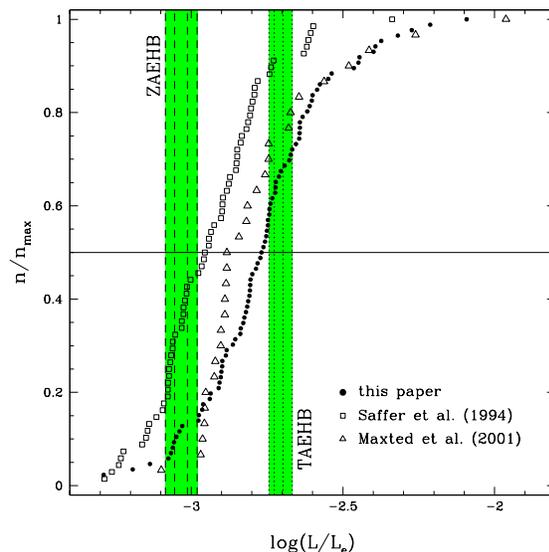}
\begin{minipage}{8.8cm}
	\caption[]{Plot of (normalized) numbers of stars versus luminosity (with respect to the Eddington
   luminosity $L_e$, see Eq. \ref{luminosity_f}).
  Additionally are plotted the results of Saffer et al. (\cite{Saf94}, open squares) 
  and Maxted et al. (\cite{Max01}, open triangles).
   The  dashed vertical lines denote the ZAEHB, and the dotted vertical lines
   the TAEHB for metallicities of [M/H]=0.00, $-$0.47, $-$1.48, and $-$2.22, respectively 
   from the left to the right (Dorman, Rood \& O'Connell \cite{Dor93}).
 	\label{logl_le_n}
}
\end{minipage}
\end{figure}
We can, however, compare the samples in a global sense using the cumulative 
luminosity functions.
In Fig. \ref{logl_le_n} we plot these functions for our sample and those of
Saffer et al. (\cite{Saf94}) and Maxted et al. (\cite{Max01}).
The luminosity is expressed in units of the Eddington luminosity (Eq. \ref{luminosity_f}).
Additionally the positions of the ZAEHB and the TAEHB are indicated. 
Since the metallicity of the stars is unknown, we have plotted models for various metallicity 
ranging from [M/H]=$-2.22$ to 0.00 (Dorman, Rood \& O'Connell \cite{Dor93}).
Note that the position of the TAEHB is model-dependent, because poorly understood processes
such as semiconvection and convective instabilities (breathing pulses, Castellani et al. \cite{Cas85})
play a role.
As can be seen from Fig. \ref{logl_le_n}, the overall shape of the
cumulative luminosity function is similar for all three samples when the considerable
smaller number of stars in the Maxted et al. (\cite{Max01}) sample is taken into
account. However, there is an offset of about 0.2 dex in luminosity between the  
Saffer et al. (\cite{Saf94}) sample and ours, whereas the relation for the 
Maxted et al. (\cite{Max01}) sample is in better agreement with ours.
Possible reasons for the discrepancy are
different observations, the different
synthetic spectra used in the analysis, or both. Maxted et al. (\cite{Max01}) and our study is based 
on the same grid of NLTE models for stars hotter than $27\,000$K and metal line blanketed 
LTE models for the cooler ones, whereas Saffer et al. (\cite{Saf94}) use metal-free LTE models 
for all stars.
The observations of Maxted et al. (\cite{Max01}) are of considerably higher spectral resolution
than that of Saffer et al. (\cite{Saf94}) and ours and on average have a better S/N than
our data. 
The wavelength coverage of the observations also varies. 
Differences in synthetic spectra calculated from different 
sets of model atmospheres have already been reported.
Heber, Reid \& Werner (\cite{Heb00})
studied the influence of NLTE effects versus metal line blanketing in LTE and found that
NLTE models give luminosities larger by about 0.1 dex than metal line blanketed LTE models.
Therefore  about half of the observed offset between the cumulative luminosity functions
of Saffer et al. (\cite{Saf94}) and ours can be traced back to NLTE effects, but an 0.1 dex 
offset remains unexplained.
We have embarked on a more detailed investigation of this problem. It will be presented in a
forthcoming paper.

Our analysis reveals an apparent correlation between the
photospheric helium content and the stellar parameters of a sdB star: The larger the
effective temperature and/or gravity, the larger the helium abundance.
However, for sdB stars, $T_{\rm eff}$ and gravity are strongly connected and a plot
of helium abundance versus luminosity does not reveal any correlations.
There is general consensus that the low helium abundance of sdB stars is due to 
diffusion processes. Simple diffusion models assume the abundances to be set by the 
equilibrium of gravitational and radiative forces. Such models predict helium abundances 
far lower than observed (Fontaine \& Chayer \cite{Fon97}).
Weak radiation-driven stellar winds, however, are likely to be present.
Calculations by  Fontaine \& Chayer (\cite{Fon97}) and Unglaub \& Bues (\cite{Ung01})
indeed show that a better agreement of the predicted helium abundance with observations
can be achived by considering mass loss rates of the order of $10^{-14}- 10^{-12} M_{\sun}$/year.
Radiation-driven wind theory predicts mass loss rates to increase with luminosity
(Pauldrach et al. \cite{Pau88}). However, no such trend becomes apparent on our 
observations.
Therefore, we conclude that other physical processes must be considered.

Additionally, a population of stars with very low helium abundances was identified 
when the helium abundance is plotted versus the effective temperature.
These stars clearly seperate from the bulk (see Fig. \ref{nhe_teff}).
The separation of these stars is much less evident when we plot
the helium abundance versus the gravity (Fig. \ref{nhe_logg}) or the luminosity 
(Fig. \ref{logl_le__loghe_h}).
This phenomenon provides evidence that surface abundances of sdB stars are not a simple
function of their position in the HR diagram. It rules out time-independent diffusion
models and points to a dependence on the star's history.
Due to the discovery of Maxted et al (\cite{Max01}), that about 2/3rd
of all sdB stars appear to be RV variable, it is very likely that  
many sdB stars in our sample are members 
of a close binary system with an unseen companion, like \object{HS\,2333+3927} already
discovered.
Aznar Cuadrado \& Jeffery (\cite{Azn02}) suggest that
short-period binaries may have a larger photospheric 
helium content than long-period binaries
due to tidal effects disturbing the diffusive separation inside sdB stars (see above)
in short-period systems more than in long-period systems.
Therefore, the separation into two sequences of
helium abundances possibly could be caused by their (yet undetected) binary nature.
\section{Conclusions}
We have presented the results of a spectral analysis of 111 sdB candidates selected
from follow-up observations of the Hamburg Quasar Survey. 
The analysis reveals 89 stars to be bona fide subdwarf B and subdwarf OB stars.
The remaining objects are 18 spectroscopic binaries containing a sdB component,
one HBB, and three apparently normal B stars.
Stellar parameters as well as radial velocities and distances have been determined.
The results are largely consistent with the results from the literature by other groups,
when NLTE effects are accounted for. To resolve the reason for the remaining differences, 
however, a detailed investigation of systematic errors caused by
different observational material and by the use of different model atmospheres is required.

Additionally, there remain two open questions:
What physical processes cause the discovered correlation of the helium abundance with
the effective temperature?
Why is there a separation into two sequences of sdB stars in the $T_{\rm eff}$--helium abundance 
plane? 
To understand these phenomena, more observations and further calculations are urgently needed.

Last but not least, our spectral analysis was also the starting point of another investigation:
The majority of our programme stars lie in a temperature range where non-radial pulsations 
have been predicted to occur (Charpinet et al. \cite{Cha96}) and have indeed been
observed (Kilkenny et al. \cite{Kil97}).
Therefore we initiated a collaboration with two groups in Norway and Italy in 1999
to search for pulsating sdB stars in our sample.
All of our stars 
will be observed for light variations.
Up to June 2002, about 70 HQS sdB stars had been observed and nine  
(\object{HS\,0039+4302}, \object{HS\,0444+0408}, \object{HS\,0702+6043}, 
\object{HS\,0815+4243}, \object{HS\,1824+5745}, \object{HS\,2149+0847}, \object{HS\,2151+0857}, 
\object{HS\,2201+2610}, and \object{HS\,2303+0152}) were found to be pulsating 
({\O}stensen et al. \cite{Ost01a}, \cite{Ost01b}, Dreizler et al. {\cite{Dre02}, Silvotti et al. \cite{Sil02}).
This represents about one pulsator in ten sdB stars.
It also means that about one third of all known sdBV stars discovered so far 
have been drawn from our investigation presented here.
The photometric monitoring also led to the discovery of a short period eclipsing binary
of the HW Vir type (\object{HS\,0705+6700}, Drechsel et al. \cite{Drec01})
which is only the third member of this class.
\begin{table*}
\caption[]{Spectral type, position and results of our spectral analysis for all programme stars.
}
\label{results}
\vspace{24cm}
\includegraphics{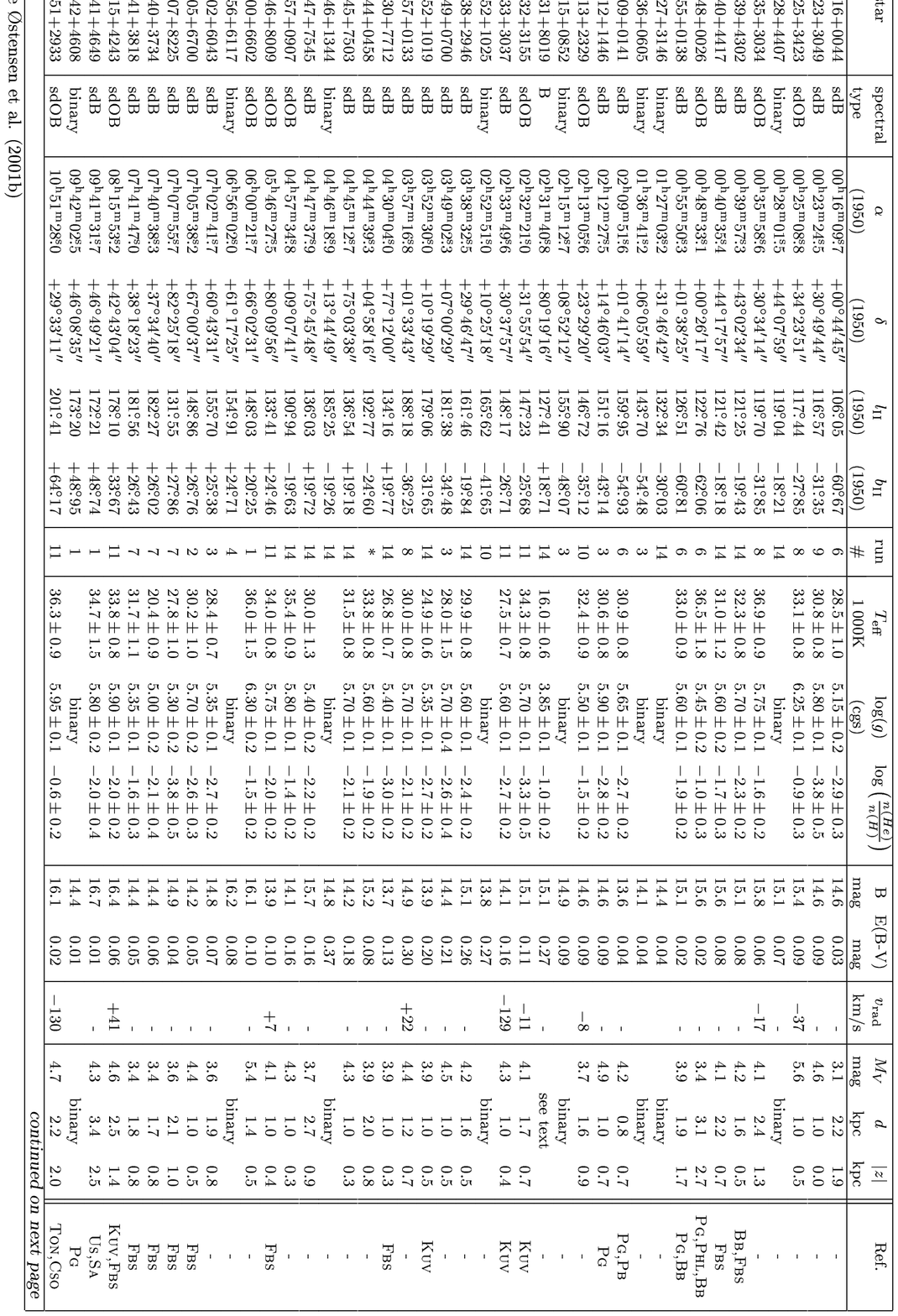}
\end{table*}
\begin{figure*}
\vspace{24cm}
\includegraphics{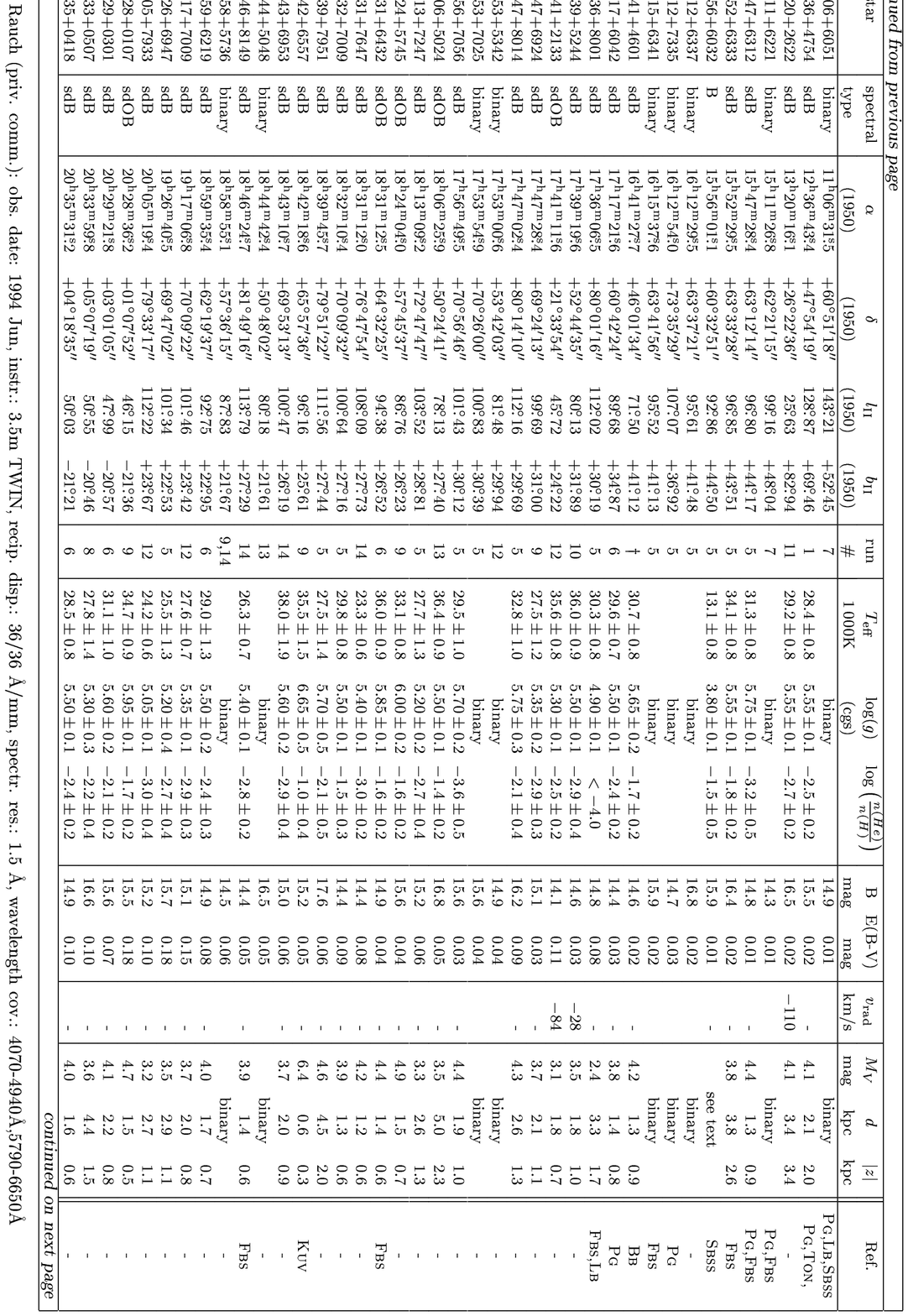}
\end{figure*}
\begin{figure*}
\vspace{24cm}
\includegraphics{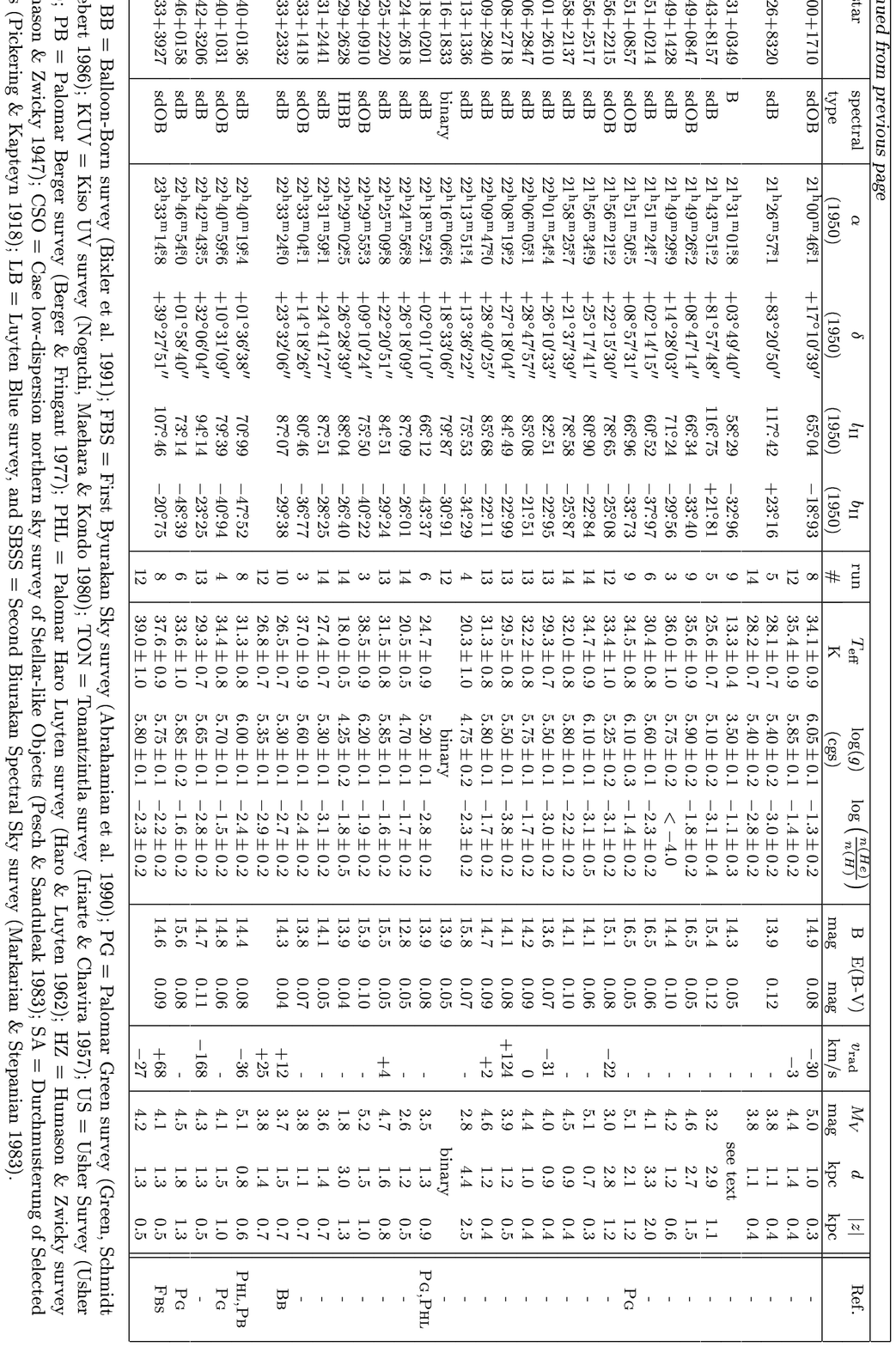}
\end{figure*}
\begin{figure*}[tbh]
\begin{minipage}[h]{17cm}
 \begin{minipage}[h]{5.5cm}
  \begin{flushleft}
   \resizebox{\hsize}{!}{\includegraphics{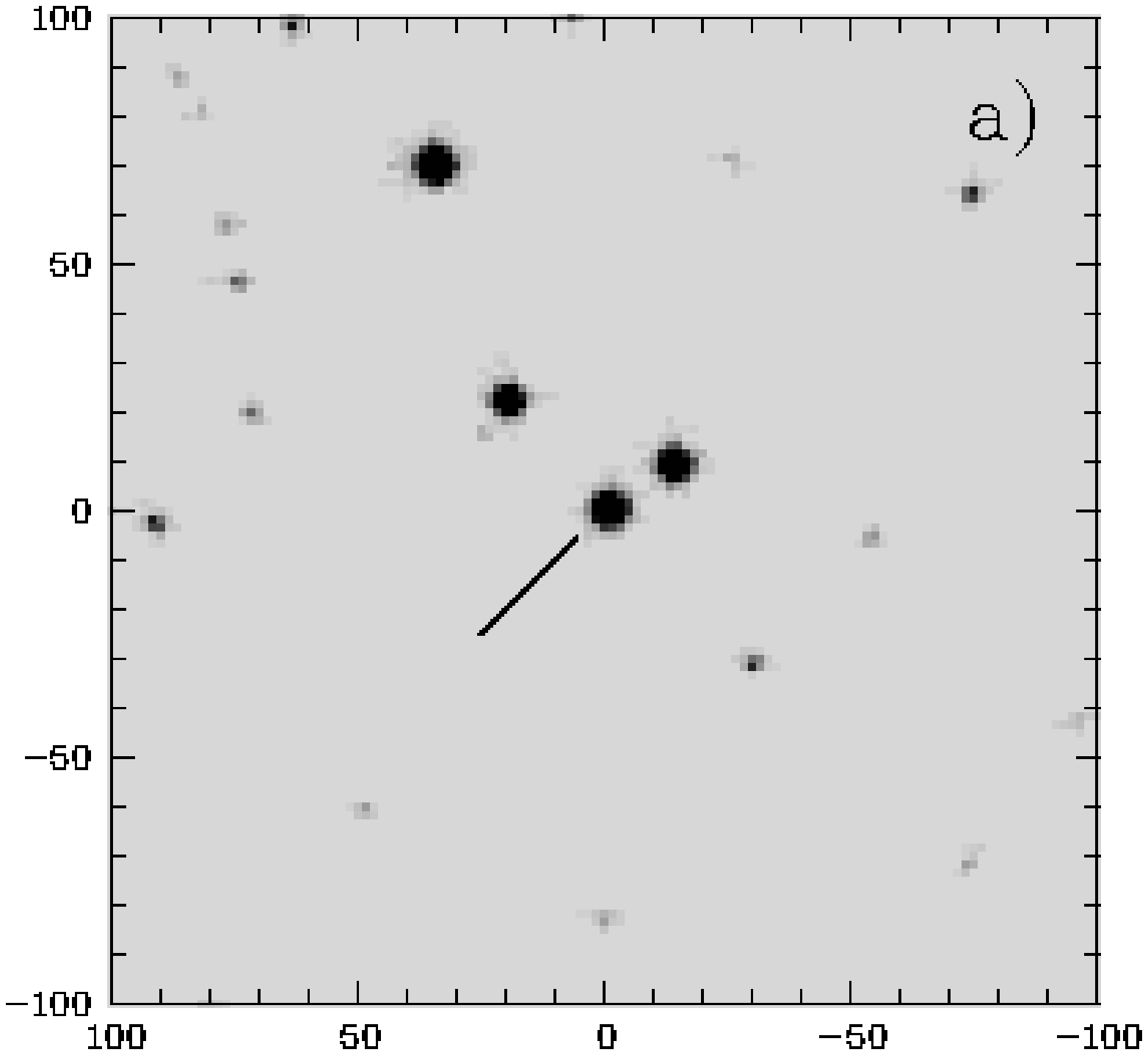}}
  \end{flushleft}
 \end{minipage}
 \hfill
 \begin{minipage}[h]{5.5cm}
  \begin{center}
   \resizebox{\hsize}{!}{\includegraphics{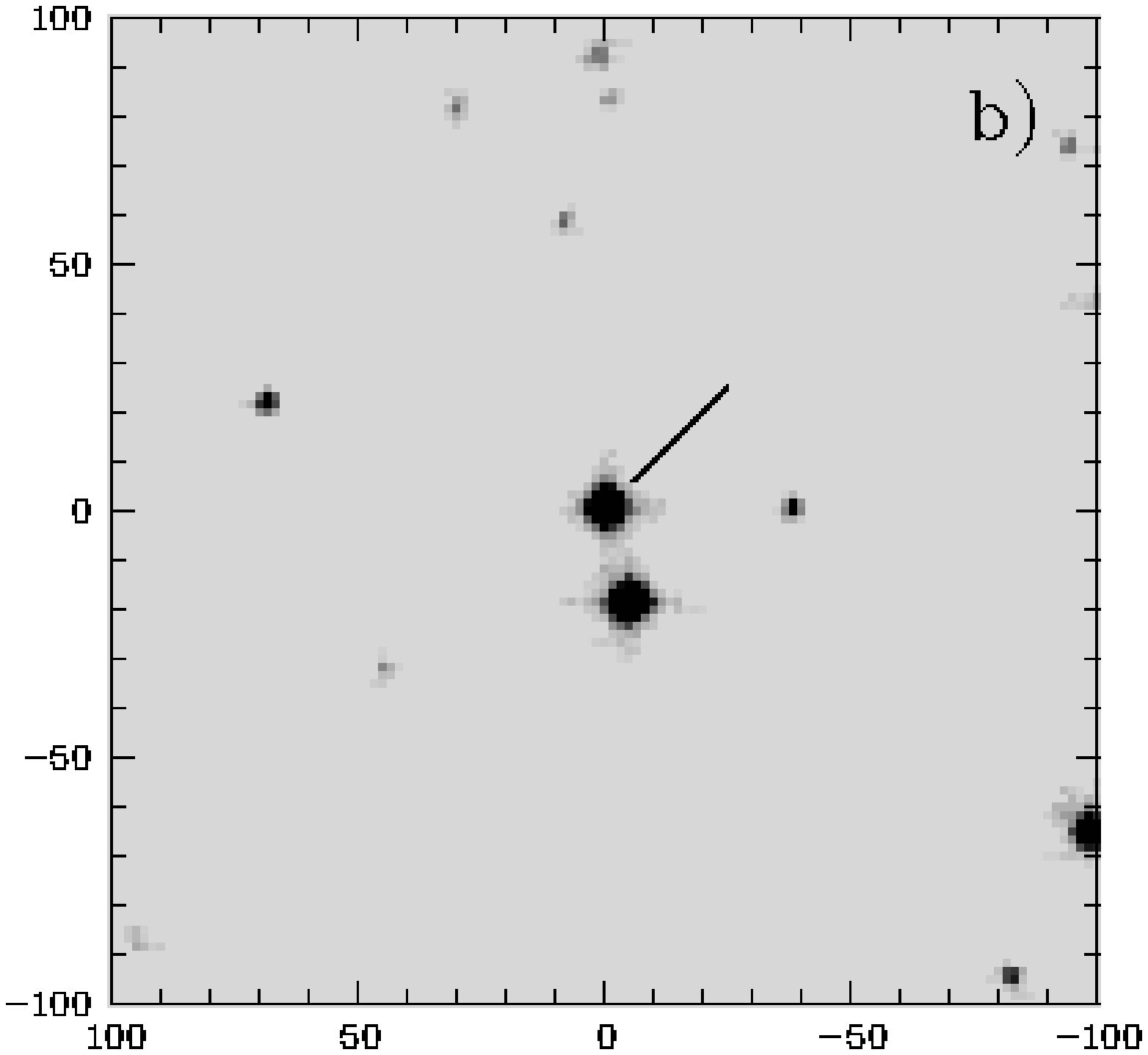}}
  \end{center}
 \end{minipage}
 \hfill
 \begin{minipage}[h]{5.5cm}
  \begin{flushright}
   \resizebox{\hsize}{!}{\includegraphics{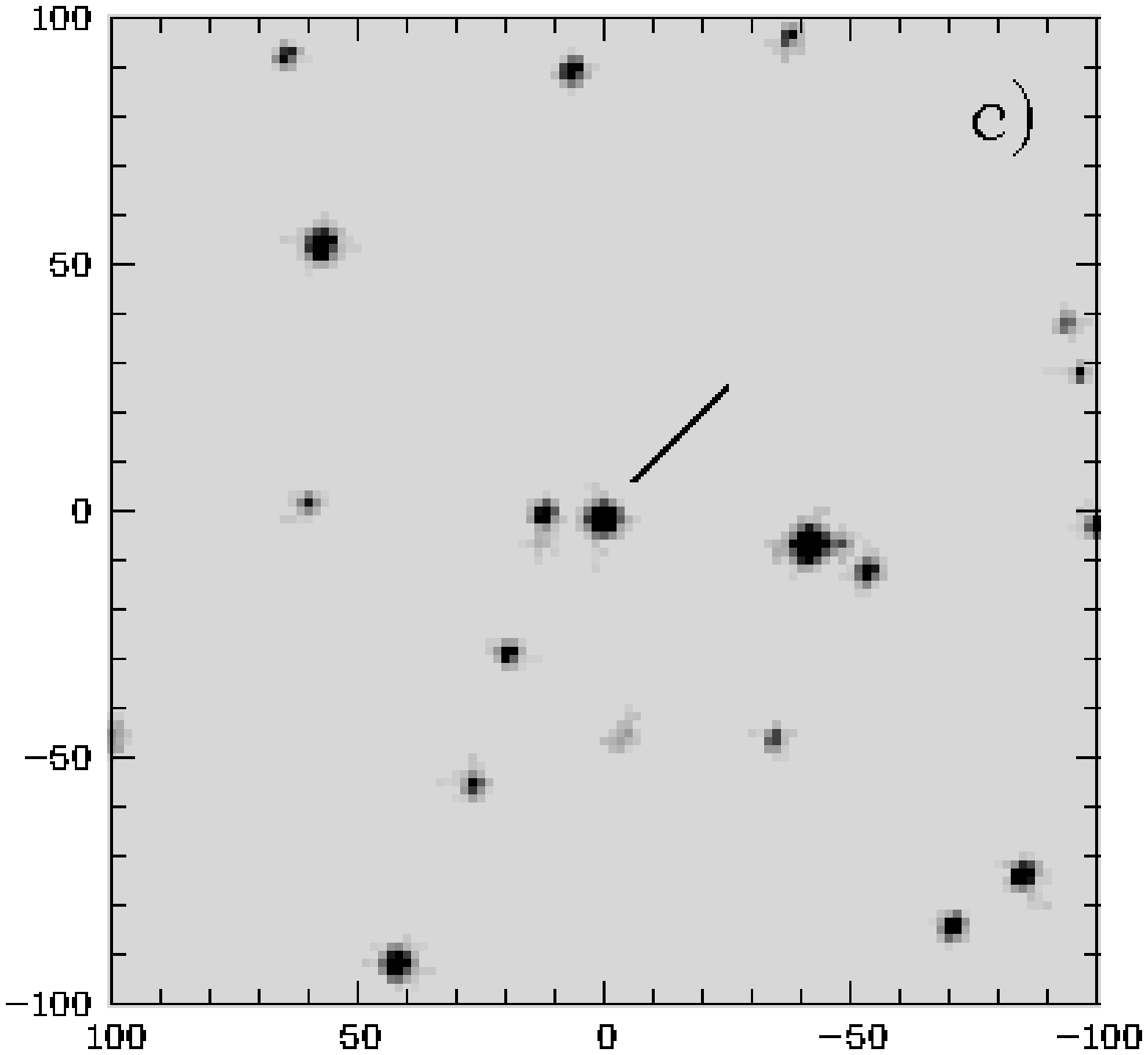}}
  \end{flushright}
 \end{minipage}
\hfill
 \begin{minipage}[h]{5.5cm}
  \begin{flushleft}
   \resizebox{\hsize}{!}{\includegraphics{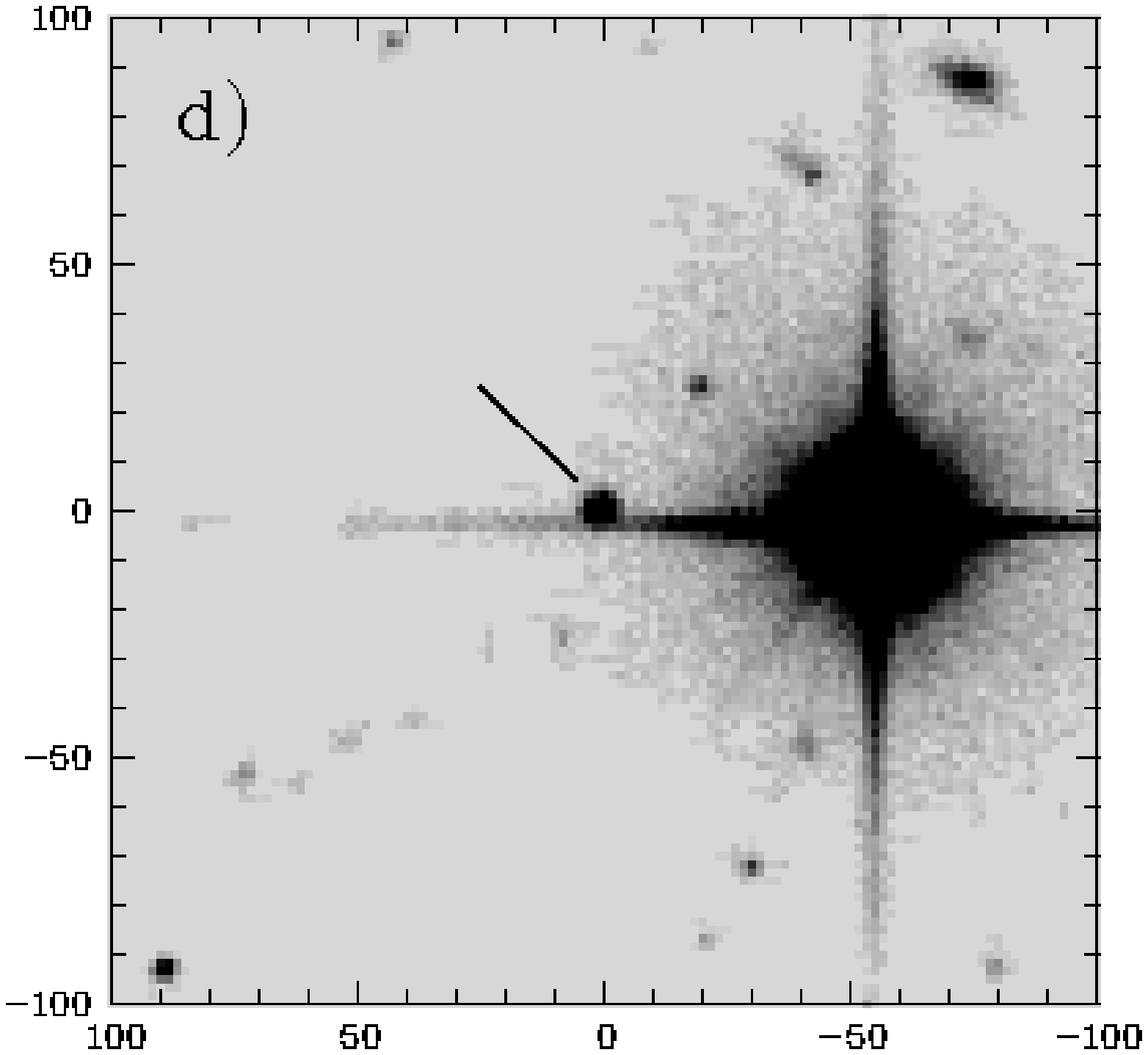}}
  \end{flushleft}
 \end{minipage}
 \hfill
 \begin{minipage}[h]{5.5cm}
  \begin{center}
   \resizebox{\hsize}{!}{\includegraphics{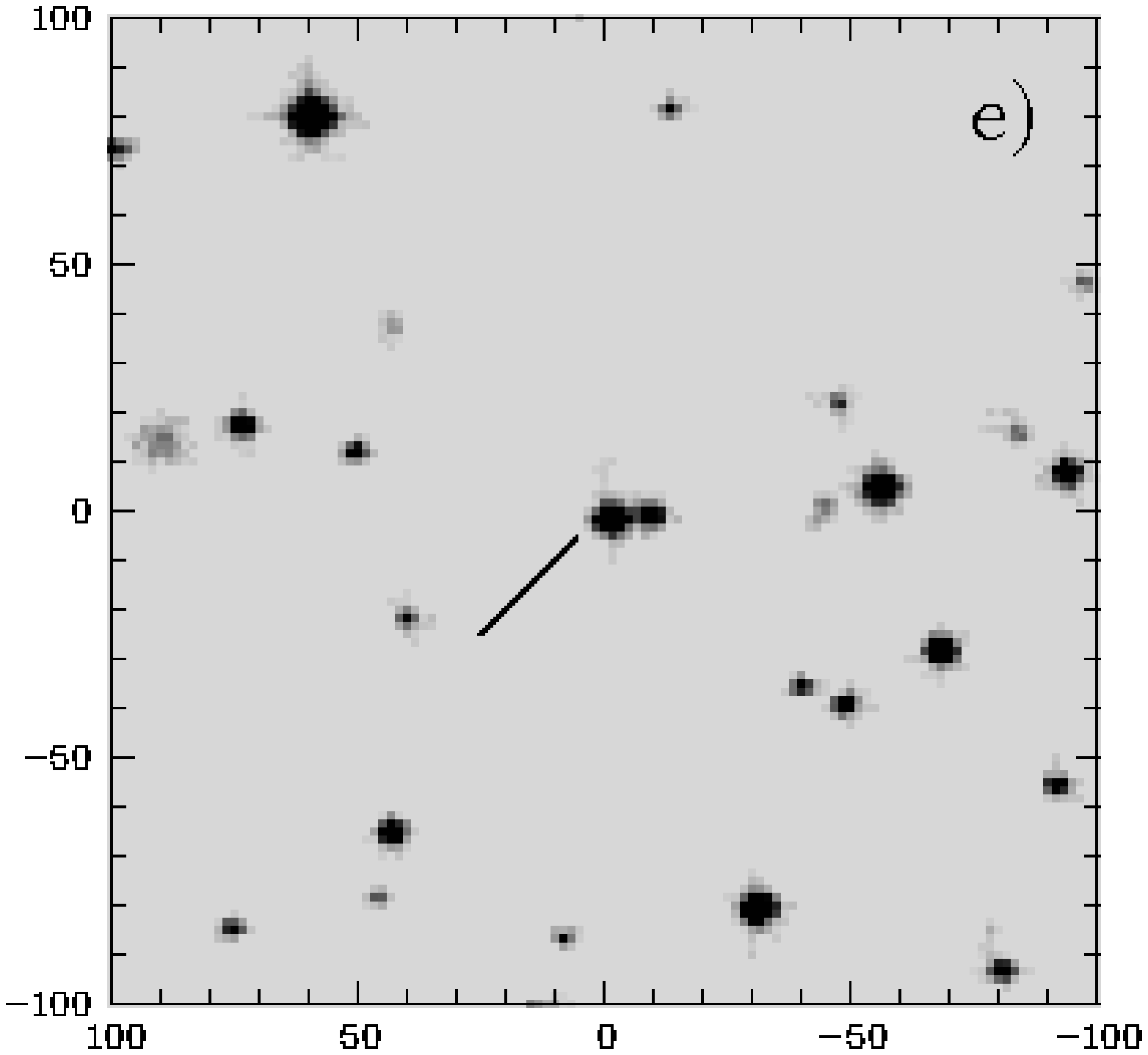}}
  \end{center}
 \end{minipage}
 \hfill
 \begin{minipage}[h]{5.5cm}
  \begin{flushright}
   \resizebox{\hsize}{!}{\includegraphics{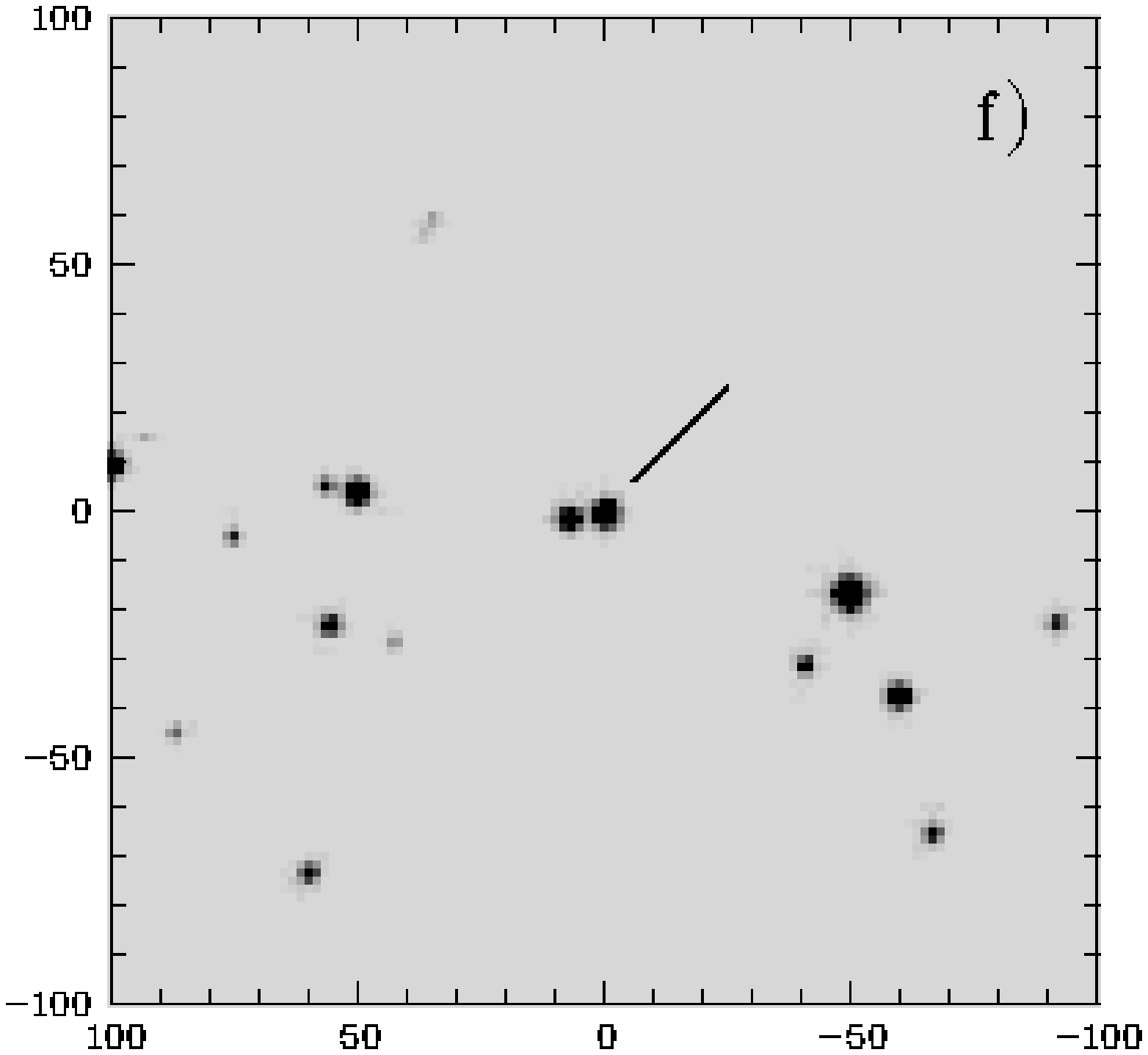}}
  \end{flushright}
 \end{minipage}
\hfill
 \begin{minipage}[h]{2.75cm}
  \begin{flushleft}

  \end{flushleft}
 \end{minipage}
 \begin{minipage}[h]{5.5cm}
  \begin{flushleft}
   \resizebox{\hsize}{!}{\includegraphics{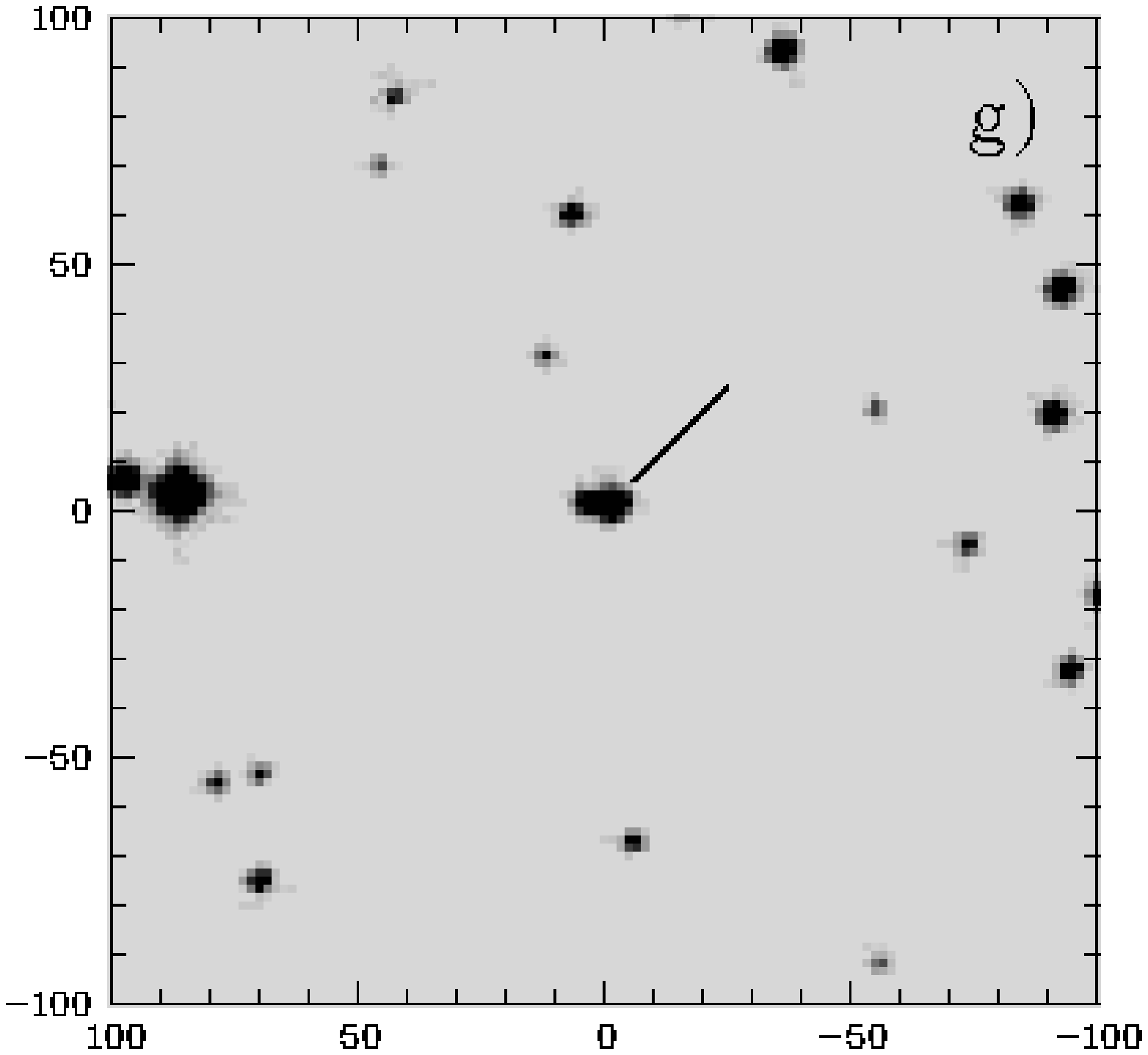}}
  \end{flushleft}
 \end{minipage}
 \hfill
 \begin{minipage}[h]{5.5cm}
  \begin{flushright}
   \resizebox{\hsize}{!}{\includegraphics{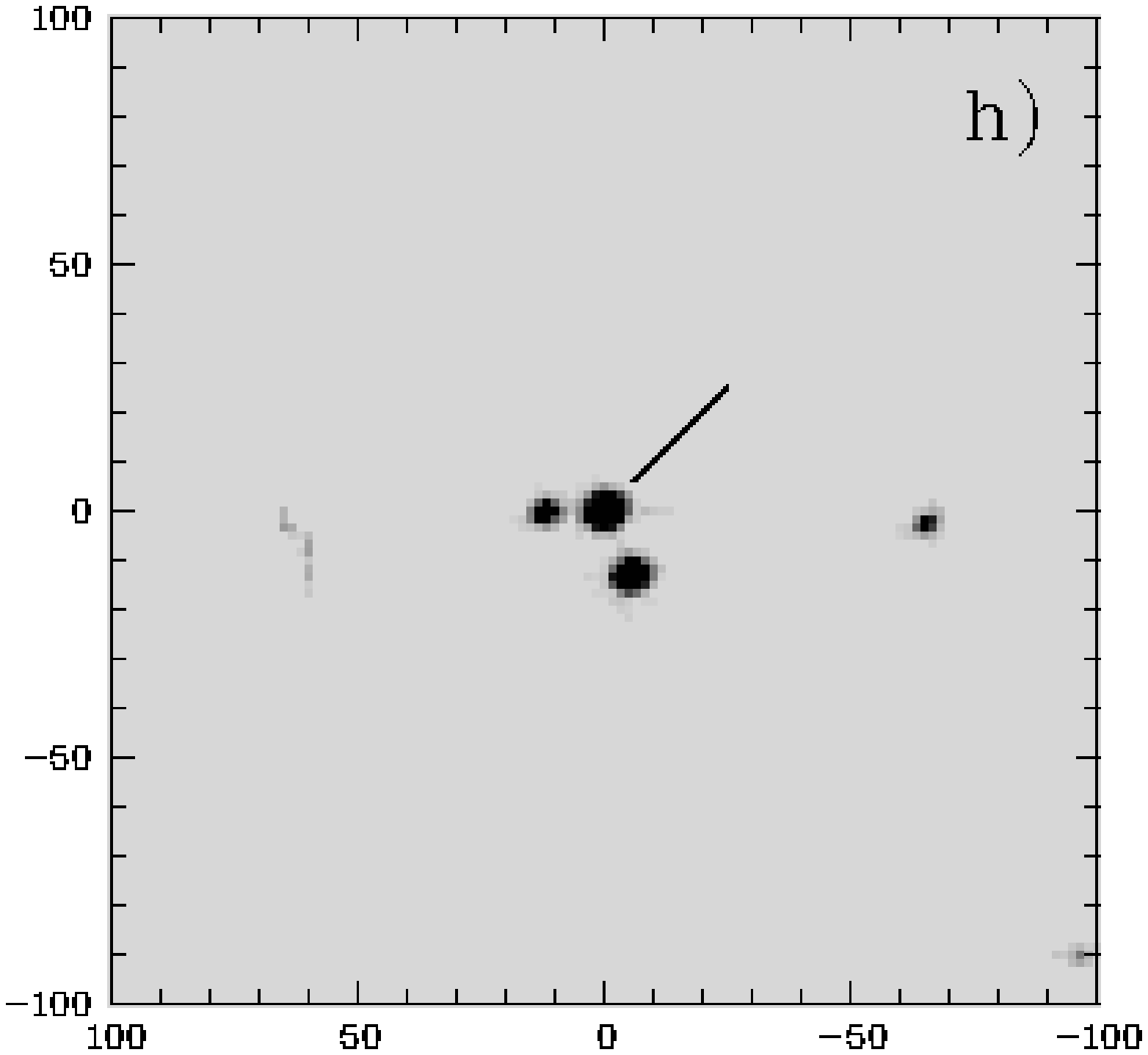}}
  \end{flushright}
 \end{minipage}
 \begin{minipage}[h]{2.75cm}
  \begin{flushright}

  \end{flushright}
 \end{minipage}
\end{minipage}
\caption[]{HQS finding charts of selected subdwarfs: a)~HS\,0039+4302, b)~HS\,0213+2329, 
c)~HS\,0600+6602, d)~HS\,1320+2622, e)~HS\,2100+1710, 
f)~HS\,2143+8157, g)~HS\,2206+2847, h)~HS\,2229+0910. The charts are
centered on the coordinates given in Table \ref{results} and the size 
is 200\arcsec $\times$ 200\arcsec. East is left and North is up.}
 	\label{findingcharts}
\end{figure*}
%
%
\begin{acknowledgements} 
Thanks go to all those colleagues who participated in the observing campaigns:
S.~Jordan, R.~M\"oller, H.~Marten, S.~Haas, 
and also to the staff of the Calar Alto observatory, Spain, 
for their valuable assistance during our visits.
Additionally we are grateful to T. Rauch for providing us with his spectrum.
We thank Dr. C. S. Jeffery, the referee, for his valuable
suggestions which helped to improve the paper. 
H.E. acknowledges financial support by the German 
research foundation DFG under grant He 1354/30$-$1 and
for several travel grants to the Calar Alto observatory.
The HQS was supported by DFG grants Re 353/11 and Re 353/22.
This research has made use of the SIMBAD database,
operated at CDS, Strasbourg, France.

\end{acknowledgements}


\end{document}